\documentclass[a4paper]{article}

\usepackage{times,amsmath}
\usepackage{txfonts}
\usepackage{graphicx}
\usepackage{subfig}
\usepackage{hyperref}
\usepackage{caption}
\usepackage{lmodern}
\usepackage{tabularx}
\usepackage{tabulary}
\usepackage{longtable}
\usepackage{footnote}
\usepackage{enumerate}
\usepackage[T1]{fontenc}
\usepackage[utf8]{inputenc}
\usepackage{authblk}

\def\XMM{\textit{XMM-Newton} }
\def\xr{\textit{X-ray} }
\def\xrs{\textit{X-rays} }
\def\cgs{$erg\,s^{-1}\,cm^{-2}\,$}

\begin{document}

\title{Analysis of X-ray spectral variability and black hole mass determination of the NLS1 galaxy Mrk 766.}

\author[1,2]{S. Giacch\`{e}}
\author[3]{R. Gilli}
\author[2]{L. Titarchuk}

\affil[1]{Max-Planck-Institut f\"ur Kernphysik, Saupfercheckweg 1, 69117 Heidelberg, Germany}
\affil[2]{Dipartimento di Fisica, Universit\`{a} di Ferrara, Via Saragat, 1, 44100 Ferrara, Italy}
\affil[3]{INAF -- Osservatorio Astronomico di Bologna, Via Ranzani 1, 40127 Bologna, Italy}
\renewcommand\Authands{ and }

%\author{
%S. Giacch\`{e}\inst{1,2},
%R. Gilli\inst{3},
%L. Titarchuk\inst{2}
%}

	\date{November 04, 2013}

%\institute{
%Max-Planck-Institut f\"ur Kernphysik, Saupfercheckweg 1, 69117 Heidelberg, Germany
%\and
%Dipartimento di Fisica, Universit\`{a} di Ferrara, Via Saragat, 1, 44100 Ferrara, Italy
%\and
%INAF -- Osservatorio Astronomico di Bologna, Via Ranzani 1, 40127 Bologna, Italy }

\maketitle

\abstract {We present an \XMM time-resolved spectral analysis of the Narrow Line Seyfert 1 galaxy Mrk 766. We analyse eight available observations of the EPIC-pn camera taken between May 2000 and June 2005 in order to investigate the \xr spectral variability as produced by changes in the mass accretion rate. The $0.2-10\,keV$ spectra are extracted in time bins longer than $3\,ks$ to have at least $3\times10^4$ net counts in each bin and then accurately trace the variations of the best fit parameters of our adopted Comptonisation spectral model. We test a bulk-motion Comptonisation (BMC) model which is in general applicable to any physical system powered by accretion onto a compact object, and assumes that soft seed photons are efficiently up-scattered via inverse Compton scattering in a hot and dense electron corona. The Comptonised spectrum has a characteristic power-law shape, whose slope was found to increase for large values of the normalisation of the seed component, that is proportional to the mass accretion rate $\dot{m}$ (in Eddington units). Our baseline spectral model also includes a warm absorber lying on the line of sight and radiation reprocessing from the accretion disk or from outflowing matter in proximity of the central compact object. Our study reveals that the normalisation-slope correlation, observed in Galactic Black Hole sources (GBHs), also holds for Mrk 766: variations of the photon index in the range $\Gamma\sim1.9-2.4$ are indeed likely to be related to the variations of $\dot{m}$, as observed in \xr binary systems. We finally applied a scaling technique based on the observed correlation to estimate the BH mass in Mrk 766. This technique is commonly and successfully applied to measure masses of GBHs, and this is the first time it is applied in detail to estimate the BH mass in an AGN. We obtain a value of $M_{BH}=1.26^{+1.00}_{-0.77}\times 10^6\,M_{\odot}$ that is in very good agreement with that estimated by the reverberation mapping.}

%\keywords{Black Hole physics -- Galaxies: active -- X-rays: general}

\section{Introduction}	\label{sec:intro}

The study of the \xr spectral properties of accreting compact objects is crucial in modern astronomy to understand the physics of the accretion process and to investigate the distinctive features of these objects. The main questions which need to be answered are whether there are remarkable differences in the accretion onto Black Holes (BHs) and Neutron Stars (NSs) in \xr Galactic Binary Systems, whether there is a unified accretion scheme involving both Galactic Black Hole sources (GBHs) and Active galactic Nuclei (AGNs) and whether, in this scheme, AGNs show the same variability patterns observed in GBHs.\\
Focusing on the accretion mechanism onto GBHs, the observed phenomenology is usually described in terms of BH state classification. A BH transient going into outburst leaves the \textit{quiescent state} and enters a \textit{low-hard state}, i.e. a low-luminosity state whose energy spectrum is dominated by a Comptonisation component combined (convolved) with a weak thermal component. The source might then evolve towards a \textit{high-soft state} or \textit{very high-soft state}, characterised by high luminosity and dominant thermal component\footnote{here we described only the main observed spectral states, for a complete review see \cite{RM06}.}. For Galactic Black Hole sources (GBHs) in \xr Binary Systems, the relationship between timing, spectral and mass accretion properties has been intensively studied in the \xr energy window \cite{ST07,ST09, ST11} with a generic Comptonisation model (BMC) which consistently convolves a black body (BB) spectrum originating from the accretion disk and the Green function of the Comptonisation process 
%\textbf{
(which is a broken power-law)
%}
 that takes place in the hot and dense electron corona (Compton Cloud, CC) surrounding the central black hole. 
%\textbf{
Despite the name (bulk motion Comptonisation) that could sound a little bit confusing, the BMC broken power-law is a generic kernel which is valid for any kind of up-scattering, both thermal or bulk motion \cite{LT99}.
%} 
According to this Comptonisation model, the spectral evolution undergone by objects powered by the accretion mechanism is driven by the variations in the mass accretion rate $\dot{m}$, expressed in units of the Eddington rate $\dot{M_E}=L_E/c^2$  \footnote{we follow the definition used in \cite{TMK97} and \cite{TZ98}. Note that this differs from the definition given in \cite{SS73}, commonly used in AGN literature, by a factor of $\eta$. Therefore in our notation $\lambda=L/L_E=\eta \dot{m}$.}. The power emitted by the accreting system via the accretion mechanism is then $L=\dot{M}c^2\eta(r)\propto \dot{m}m\eta(r)$, where $m$ is the mass of the compact object in solar mass units and $\eta(r)$ is the radiative efficiency at distance r to it. The increase of $\dot{m}$ implies a rise of the soft photon supply from the innermost regions of the accretion disk (emitting BB-like radiation) that efficiently cools down and shrinks the CC via Compton scattering, softening in turn the resulting spectrum. For $\dot{m}\gg1$ the CC progressively shrinks, as electrons become very cool, until a full bulk-motion regime is established (only in black hole sources, where the bulk inflow is not hampered by a strong radiation-pressure force as in Neutron Stars). Furthermore, when the mass accretion rate decreases, the CC puffes out, a larger fraction of seed photons is up-scattered to higher energies via thermal Comptonisation and the spectrum becomes harder. These considerations are the basis to interpret the correlation between slope and normalisation $\Gamma-N_{BMC}$ observed in GBHs, where $\Gamma$ is the intrinsic photon index of the emitted power-law and $N_{BMC}$ is the normalisation of the BMC model. In particular for BH sources, the correlation sometimes shows a $plateau$ at low values of both $\Gamma$ and accretion rate. For higher values of $\dot{m}$, a slope becomes steeper which is dictated by the soft photons cooling efficiency when $\dot{m}$ increases (\cite{TS09,ST10}). Thus the spectral transition from the \textit{low-hard state} to the \textit{high-soft state} takes place, which is followed by a saturation of the photon index for large values of $\dot{m}$. In fact, the photon index is an inverse of the Comptonisation parameter $Y=\kappa N_{sc}$ that describes the efficiency of the Compton scattering, where $\kappa$ is the mean energy gain per scattering and $N_{sc}$ is the number of scattering events ($\kappa=4kT/m_ec^2$ for normal thermal Comptonisation). \cite{ST09} showed that when the bulk-motion is established $\kappa\propto1/\tau$ and $N_{sc}\propto \tau$, where $\tau$ is the optical depth. Because of this the Comptonisation parameter becomes constant and so does $\alpha\propto1/Y$ ($\Gamma=\alpha+1$, see equation B7 in \cite{ST09}, hereafter ST09, for details). The final saturation (\textit{very high-soft state}) usually occurs in the range $\Gamma\sim 2.1-3$ depending on the contribution to the final spectrum of both the thermal Comptonisation and the bulk-motion Comptonisation: the larger $\Gamma$, the more dominant is the bulk-motion effect with respect to the thermal Comptonisation.\\
%\textbf{
A secondary effect predicted by the BMC theory is an high energy cut-off due to Compton recoil effect, that for GBHs has been observed in the range $50-250$ keV \cite{ShT10}.
%}\\
Given that the bulk-motion and the related saturation of the photon index can only establish in presence of the event horizon of black hole sources, the saturation $plateau$ in the correlation pattern is a conclusive piece of evidence on the nature of the central compact object. The photon index saturation has been observed for example for the GBHs Cyg X-1, XTE J1550-564 and XTE J1650-500 
%\cite{ST09}.
(ST09).\\

In the spirit of the grand unification scheme between GBHs and the Super Massive Black Holes (SMBHs) likely powering Active Galactic Nuclei (AGNs), we present the first detailed application of the BMC model to an extragalactic BH. \cite{GT11} attempted to perform this analysis on a sample of AGNs with a single \xr observation. The final purpose of our analysis is instead to use a long \xr monitoring of an extragalactic source to verify whether the high variability, shown by AGNs on every observed time-scale (down to few hundreds seconds), is actually driven by changes in $\dot{m}$. In addition, we aim to perform an estimate of the mass of the SMBH powering an AGN measuring the $\Gamma-N_{BMC}$ correlation and scaling it to the same relation obtained for a reference GBH.\\
In order to carry out this analysis we selected Mrk 766, a low mass ($M_{BH}\sim1.8\times 10^6\,M_{\odot}$, \cite{B09}) Narrow Line Seyfert 1 galaxy that has undergone an intense \xr monitoring by the \XMM observatory from May 2000 to June 2005, with eight observations lasting at least $30$ ks giving an overall monitoring lasting $\sim711$ks. NLS1 galaxies are usually distinguished because of their rapid optical and \xr variability, steep \xr spectra and mass accretion rates close to the Eddington value \cite{K08}. Thus, they constitute a perfect target to sample a variety of spectral states and possibly see the saturation of the intrinsic photon index. In addition, the large exposure time and the high throughput in the $0.2-10$ keV energy band of \XMM provide a good photon statistics to well constrain the spectral parameters of the applied models. This is crucial to perform a consistent analysis given the complexity of the AGNs spectra. Usually, for GBHs the BMC model is able to account for almost the whole 
$1-100$ keV spectrum. However, the power-law description of AGN spectra is only good to the first order, since these sources usually show additional components in the $0.2-10$ keV energy band, such as a soft excess, a spectral hardening at $E\gtrsim5$ keV (the so-called \textit{Compton Hump}), as well as absorption and emission features.\\
The aim of the present work is to sample the intrinsic \xr spectral changes in order to populate the $\Gamma-N_{BMC}$ correlation. 
%\textbf{
{\it Given the limited band available with \XMM, it will not be possible to study the high energy cut-off expected in the BMC scenario and thus we will just neglect the behavior of this spectral feature in the present work}.
%}\\
The paper is organized as follows. The \XMM observations and the analysis of the light curves, namely the count rate in specific energy bands as a function of time, are described in Sect. \ref{sec:pre}, Sect. \ref{sec:spec} deals with the criteria that led to the composition of the time intervals and with the subsequent time resolved spectral analysis that we performed on our sample. In Sect. \ref{sec:disc} we discuss the results obtained with the baseline spectral model and we present the results of the scaling of the mass of Mrk 766, while in Sect. \ref{sec:conc} we draw our conclusions.

\section{Data reduction and preliminary analysis}	\label{sec:pre}

We only consider EPIC-pn data. All observations were performed in small-window mode. The EPIC-pn data reduction was performed using the Science Analysis Software (SAS) version 11.0, following the standard pipeline suggested by the \XMM Science Operation Centre (SOC) in the SAS 'threads'\footnote{http://xmm.esac.esa.int/sas/current/documentation/threads/}.
\begin{table}[h]
\caption{\XMM observation log of Mrk 766.}
\begin{center}
\begin{tabular}{cccc}
\hline
\hline
revolution & starting time & total & net \\
number & [yyyy-mm-dd] & exposure [s] & exposure [s]\\
\hline
0082 & 2000-05-20 & 58835 & 25540\\
0265 & 2001-05-20 & 129906 & 88170\\
0999 & 2005-05-23 & 95510 & 65274\\
1000 & 2005-05-25 & 98910 & 58882\\
1001 & 2005-05-27 & 98918 & 59179\\
1002 & 2005-05-29 & 95514 & 61415\\
1003 & 2005-05-31 & 98918 & 50952\\
1004 & 2005-06-03 & 35017 & 20090\\
\hline
\end{tabular}
\end{center}
\caption*{The net exposure time already accounts for the corrections due to the detector live time (71\% of the total frame time in the small-window mode) and to the filtering of flaring background periods.}
\label{tab:obs}
\end{table}

\begin{figure*}[ht!]
\begin{center}
\subfloat
{\includegraphics[width=0.55\textwidth,angle=270]{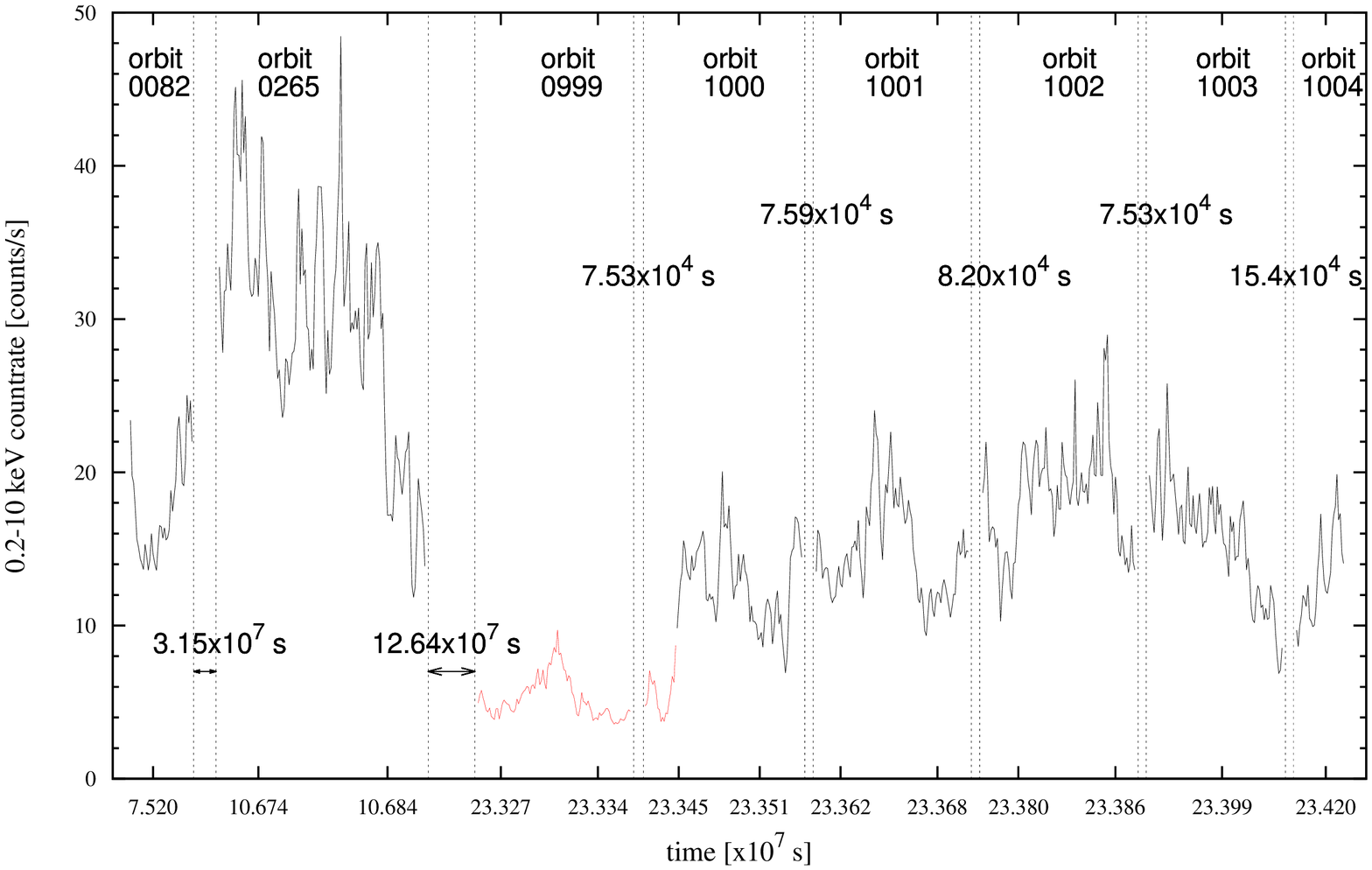}}

\subfloat
{\includegraphics[width=0.55\textwidth,angle=270]{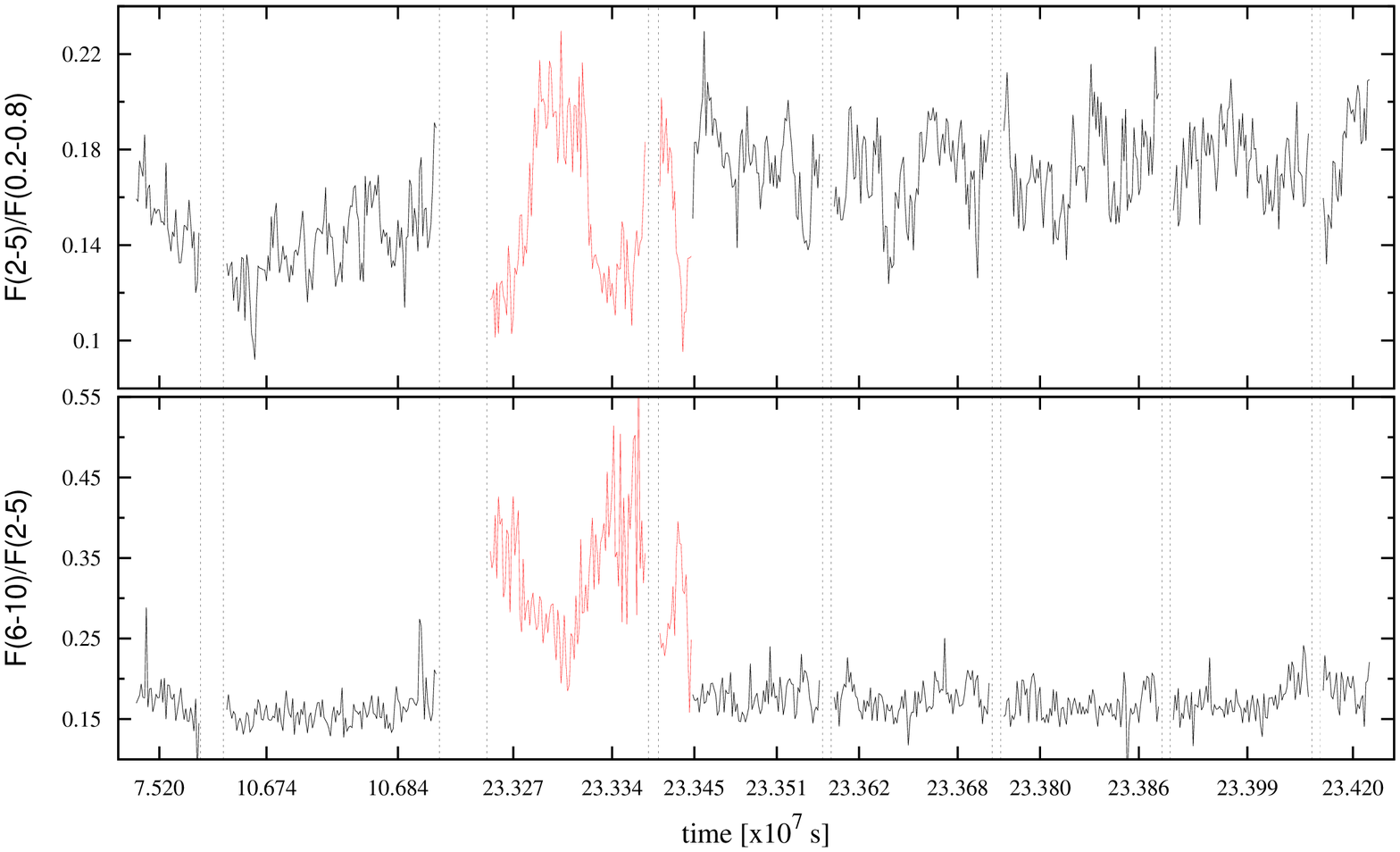}}
\caption{Light curve (upper panel) and hardness-ratio light curves (lower panel) extracted from the whole \XMM observation set. The time bin-size in both panels is $1000\,s$ to make the plots clearer. The vertical dotted lines and the related labels indicate the time gap between one observation and the next one. %\textbf{
We plot in red the data corresponding to the occultation events presented in \cite{R11} (see text for discussion).
%} 
Note that the time is measured from the beginning of the \XMM mission.}
\label{fig:lc-tot}
\end{center}
\end{figure*}
The pipeline comprises several steps, from the filtering from flaring particles background to the generation of the spectra. The flaring particles background (in particular protons with $E_p\lesssim100$ keV) affects the observation towards the end of the orbit, when the satellite approaches the radiation belt. Such kind of noise is usually very small ($\sim2\times 10^{-2}counts/s$) compared with the average source count rate that is $\sim20\,counts/s$, but a sudden bump usually appears at the end of the runs, reaching also $\sim10\,counts/s$. In order to discard time intervals affected by high flaring levels we set a threshold at the suggested value of $0.4\,counts/s$.\\
Once the event list has been filtered from flares, we selected the extraction radius of the light curve that maximizes the signal-to-noise ratio (S/N). This procedure entailed in each observation an extraction radius of $\sim40\,arcsec$, that is slightly larger than the one chosen by other authors for Mrk 766 (for example see \cite{R11}). By applying the SAS task 'epatplot', we verified that none of our observations has to be corrected for pile-up problems, not even those performed during orbit 0256 (see Fig. \ref{fig:lc-tot}), when the count rate reached the maximum value of $48\,counts/s$.\\
We extracted the light curve in three different energy ranges: the $0.2-0.8$ keV range (soft band), the $2-5$ keV range (medium band) and the $6-10$ keV range (hard band). This separation is useful to study the behavior of the different spectral components in connection with the others. In fact, the soft band flux, hereafter F(0.2-0.8), should be related to the thermal emission of the accretion disk and the medium band flux, hereafter F(2-5), should be dominated by the power-law component arising from the soft-photons up-scattered in the hot corona, while the hard band flux, hereafter F(6-10), is in principle related to the radiation fraction reprocessed in the AGN environment. The study of these three components revealed that the larger contribution to the total $0.2-10$ keV count rate, hereafter F(0.2-10), always comes from the soft band, that is usually almost one order of magnitude larger than the contribution of the medium band. In particular, the average values of the count rate for the specific case of orbit 1001 (plotted in Fig. \ref{fig:lc-tot}) are $10-12\,counts/s$ for F(0.2-0.8), $1-2\,counts/s$ for F(2-5) and $0.2-0.3\,counts/s$ for F(6-10), resulting in $17-18\,counts/s$ in the total F(0.2-10).\\
Following the procedure presented in ST09,
%\cite{ST09}, 
we studied the hardness-ratio light curves F(2-5)/F(0.2-0.8) and F(6-10)/F(2-5). As can be seen in Fig. \ref{fig:lc-tot}, Mrk 766 underwent remarkable luminosity variations on a time-scale of few hundreds seconds, accompanied by small spectral oscillations around an almost stable state on comparable times. Nevertheless, during orbit 0082, as the luminosity increases from $\sim14\,counts/s$ to $\sim24\,counts/s$, we notice an overall softening of the spectrum, implying a total variation of the order of 30\% of F(2-5)/F(0.2-0.8), while the change in F(6-10)/F(2-5) is less evident. A similar behavior, even of smaller strength, can be observed during orbit 0265 run in 2001, but in the opposite direction: an overall decrease of the luminosity, that passed from $\sim48\,counts/s$ to $\sim15\,counts/s$ in $\sim105\,ks$, is accompanied by an overall spectral hardening visible in the F(2-5)/F(0.2-0.8) hardness-ratio light curve.\\
The situation is quite different during the observations performed between May and June 2005. The luminosity variations mainly occur at almost constant spectral shape. From the lower panel in Fig. \ref{fig:lc-tot}, we notice that both the 'medium-to-soft' hardness-ratio and the 'hard-to-medium' hardness-ratio usually oscillate around $\sim0.18$. Nevertheless, we noticed the sudden drops of F(2-5)/F(0.2-0.8) down to $\sim0.12$ and the unusual peaks of F(6-10)/F(2-5) up to $\sim0.46$. It is worth noting that the most remarkable drops and peaks are related to the lowest luminosity states observed in orbits 0999 and 1000.

\subsection{Occultation episodes}	\label{sec:occ}

The low luminosity episodes, combined with the behaviour of the hardness-ratio light curves, seem to indicate a lack of photons in the $2-5\,keV$ range lasting about 20\% of the \XMM observing time. It is likely that this phenomenon is due to Broad Line Region (BLR) clouds crossing our line of sight \cite{R11}. The resulting behaviour of the hardness-ratio light curves 
%\textbf{
(see red curves in Fig. \ref{fig:lc-tot})
%}
 is probably caused by the photo-electric absorption occurring in these clouds, that produces a steepening of the observed spectral shape in the soft band.\\
Since this work is not devoted to the study of the occultation episodes, we limited ourselves to consider this possible scenario and to neglect the time intervals affected by the eclipses in the study of the $\Gamma-N_{BMC}$ correlation.

\section{Time resolved spectral analysis}	\label{sec:spec}

\subsection{Spectral sampling}  \label{sec:samp}

The study of the hardness-ratio light curves is crucial in determining the time intervals in which the spectra must be extracted. The acquisition of one single spectrum integrated over each observation, as it has been done in \cite{GT11}, would have basically averaged several different spectral states. On the other hand, this analysis revealed that Mrk 766 have passed through different states during the \XMM observations: during orbits 0082 and 0265 it underwent overall and slow intrinsic softening and hardening episodes respectively, while during 2005 the luminosity changes occurred at almost constant spectral shape, besides the occultation episodes that only caused the variation of the observed spectrum. The distinction between intrinsic and observed spectral slope is crucial, in fact for our study we are only interested in spectral variations that involve the radiation produced in the core of the system. Given the fact that it can be cumbersome to provide a correct and unambiguous parametrization of these episodes (see \cite{R11} for more details) we decided to neglect the BLR clouds eclipses.\\
Thus, we split the whole useful observing time ($\sim430\,ks$) in different intervals, trying first of all to sample the overall hardness variations observed in 2000 and 2001, and then to collect as many different spectral states as we could from the spectral oscillations in 2005. In addition, we aimed to separate the good time intervals from those affected by the occultations episodes. Since the spectral analysis involves models with several free parameters, we need a good photon statistics in each time slice to obtain restrained relative errors (corresponding to the 90\% confidence level). Given the brightness of the source, $F_{[2-10\,keV]}\sim10^{-11}\,$\cgs, the previous criterion is matched with a minimum duration of $3\,ks$, that entails a minimum of $3\times 10^4$ photons in each time interval, the only exception being the final slice extracted from orbit 1003. The end of orbit 1003 is probably affected by a BLR cloud occultation, that we tried to separate from the rest of the observation creating a dedicated time slice. Unfortunately, simultaneously to the eclipse, there is a strong background flare reaching $\sim1\,count/s$, that in that specific time interval constituted more than 10\% of the total count rate, so we excluded this bad time interval from the subsequent analysis (time slice 47 in Table \ref{tab:slices} in the Appendix).\\
The procedure we applied led us to the composition of forty-nine time intervals (fifty including slice 47), at least three from each observation, in which the spectrum was extracted. The re-binning of the spectra was performed in order for each energy channel to contain at least twenty photons: this allowed us to use the $\chi^2$ statistics to fit the models to the data and to estimate the goodness of the fits. The duration and the total counts of each time interval are collected in Table \ref{tab:slices} in the Appendix. The time resolved spectral analysis consisted in a step-by-step procedure, from the simplest to the most complicated model, and was performed with the XSPEC fitting package \cite{Xspec}, version 12.7.1.

\subsection{2-10 keV power-law fit}

\begin{figure}[h!]
\begin{center}
\includegraphics[width=0.35\textwidth,angle=270]{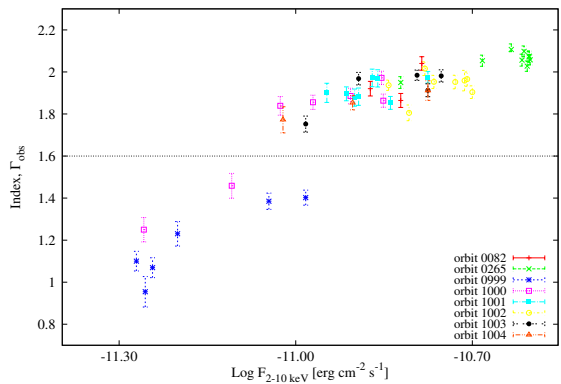}
\caption{Observed photon index $\Gamma_{obs}$ plotted against the flux in the $2-10$ keV range. Notice that the slices with the lower flux, corresponding to the occultation events (20\% of the \XMM observing time) are also characterised by the flatter power-law $\Gamma_{obs}\lesssim1.6$.}
\label{fig:po}
\end{center}
\end{figure}

First of all we fit each time slice with a simple power-law in the $2-10$ keV range to compare the observed photon index $\Gamma_{obs}$ with the photon index measured with other \xr missions. Here we refer to $\Gamma$ as observed because the power-law basically average all the spectral features across the energy range and moreover it does not account for the very first part of the non-thermal emission. In this specific energy range this simple description is on average good, with reduced chi-square $\chi^2_\nu\sim1$. As shown in Fig. \ref{fig:po}, the $\Gamma_{obs}$ variation between 0.95 and 2.1 and the flux variation in the $2-10$ keV range correlate with each other in the flux interval between $F_{[2-10~{\rm keV}]}\sim5.3\times 10^{-12}$\cgs and $F_{[2-10~{\rm keV}]}\sim2.51\times 10^{-11}$\cgs.
It is worth noting that the photon index becomes larger and larger as the flux increases. The upper part of the plot, i.e. for $\Gamma_{obs}>1.6$, is in good agreement with the measures of the photon index performed in other periods. In fact, \cite{L96} found $\Gamma_{obs}\sim1.6-2.0$ with ASCA (1993), \cite{M00} found $\Gamma_{obs}\sim2.01-2.25$ with {\it Beppo}SAX (1997) and \cite{SP09}, hereafter SP09, found $\Gamma_{obs}\sim1.6-2.5$ and $F_{[2-10~{\rm keV}]}\sim10^{-11}-6\times 10^{-11}$\cgs with {\it RXTE} (2001-2008). In particular the upper part of the plot of Fig. \ref{fig:po} constitutes the low flux tail of the trend found in SP09.\\
On the other hand, the points related to the low-luminosity episodes during 2005 \XMM observations correspond to the occultations described in Sect. \ref{sec:occ} and differ from the average behaviour. In particular, these eight time slices are characterised by low fluxes, consistent with $F_{[2-10~{\rm keV}]}\sim10^{-12}$\cgs, and by flat power-law slopes $\Gamma_{obs}\lesssim 1.5$. The fact that $\Gamma_{obs}<1.6$ has never been observed before in Mrk 766 seems to confirm that in the related intervals some unusual mechanism is at work.

\subsection{BMC model fit}	\label{sec:bmc}

\begin{figure*}[t!]
\begin{center}
{\includegraphics[width=0.4\textwidth,angle=270]{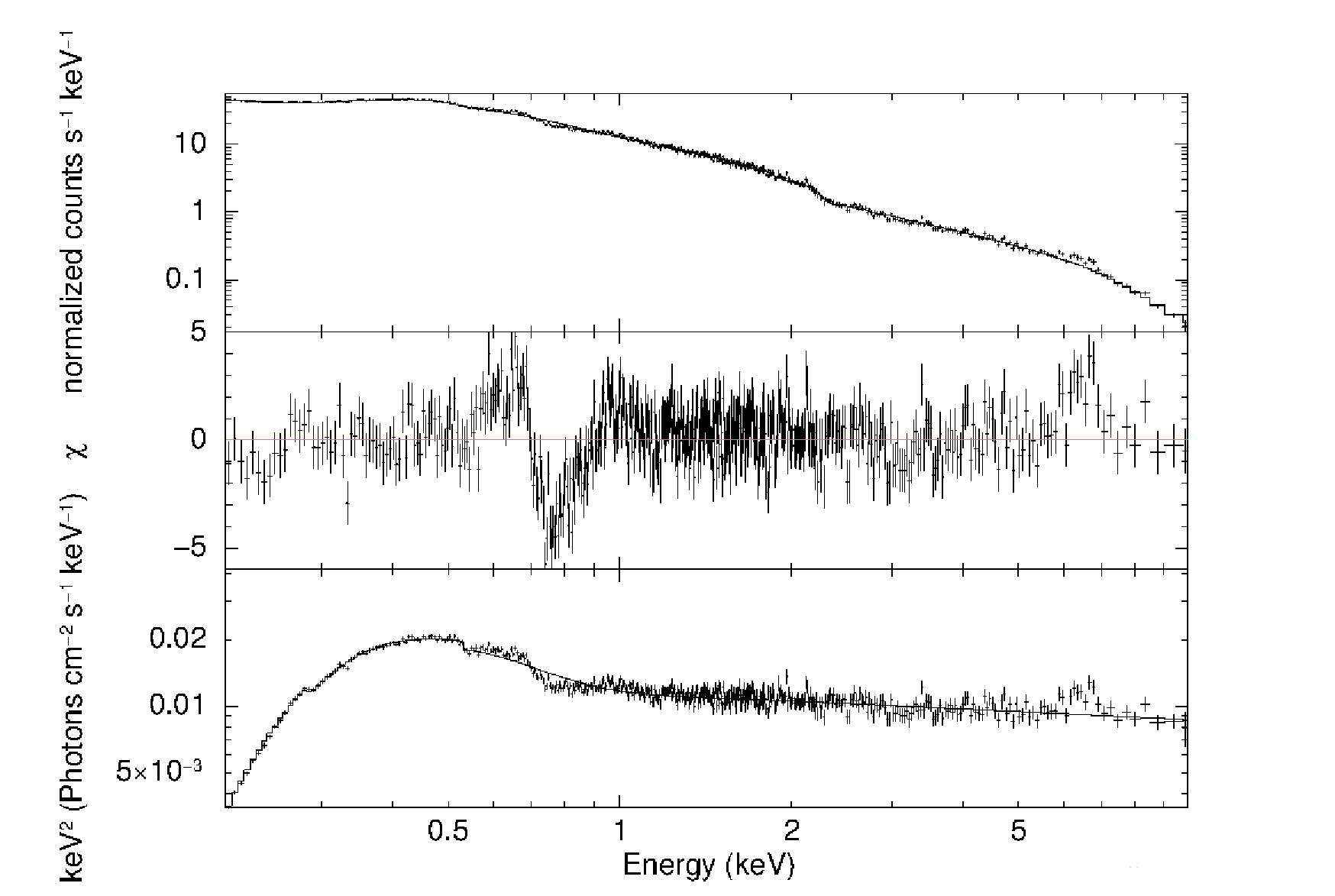}}

{\includegraphics[width=0.4\textwidth,angle=270]{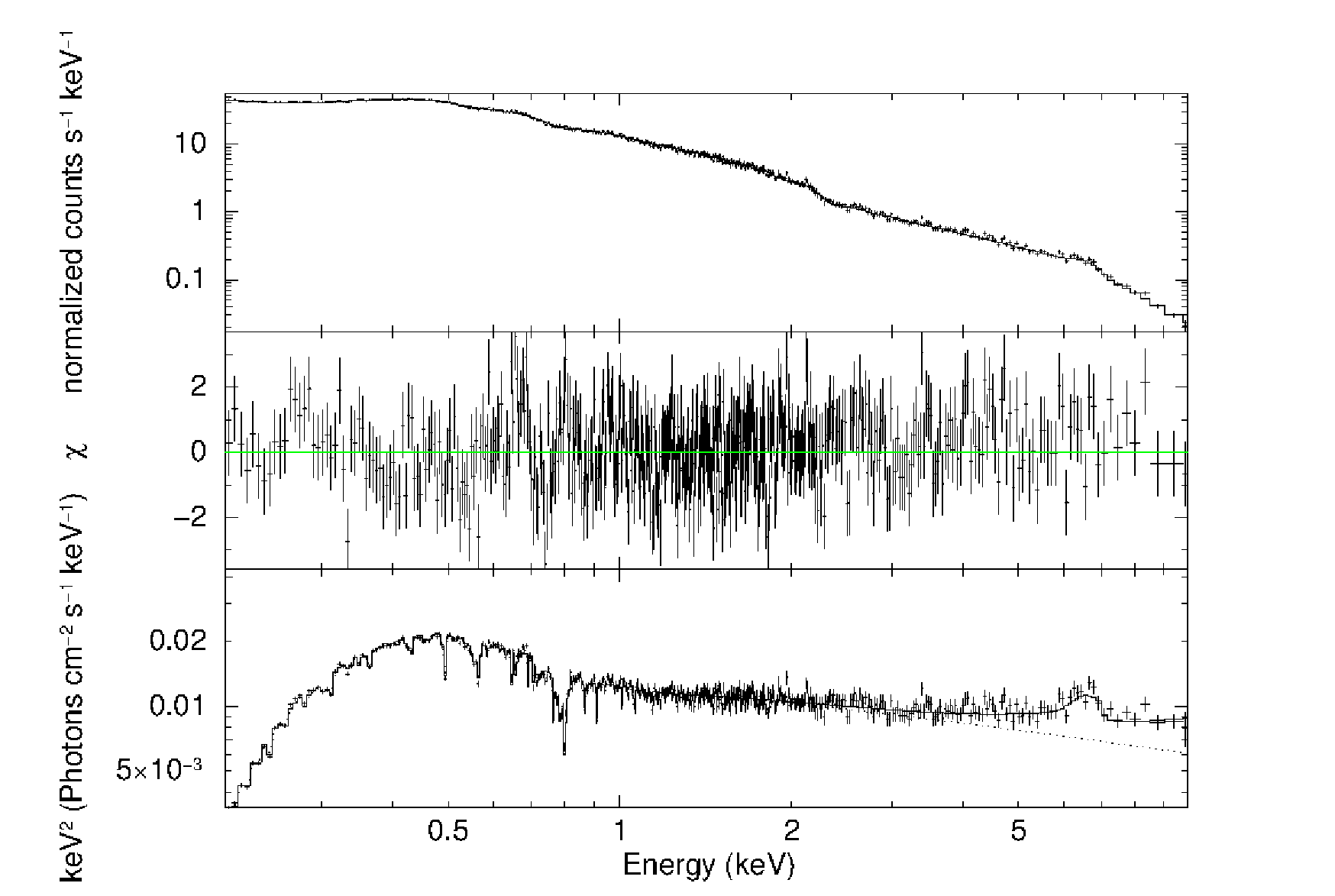}}
\caption{Data, residuals and unfolded spectra in the $0.2-10$ keV range extracted from time slice 5 in orbit 0265 in 2001 (see Table \ref{tab:abs} in the Appendix). Upper panel: the fit model comprises just the BMC and the galactic absorption. Lower panel: the fit model is improved by the addition of the PEXRAV and the ZXIPCF components. A Gaussian emission line is also included whose parameters are $E=6.59\pm0.19$ keV, $\sigma=0.33_{-0.16}^{+0.22}$ keV.}
\label{fig:spectrum}
\end{center}
\end{figure*}

Next we applied the BMC model \cite{TMK97,TZ98} in the full $0.2-10$ keV range to the forty-one time slices unaffected by the occultation events. All the parameters of the BMC model were left free to vary in all phases of our spectral analysis. \textit{It is worth noting that even though the BMC theory predicts the high energy cut-off due to recoil effect, the BMC model implemented in XSPEC does not include any parameter accounting for this phenomenon. In this sense, it can be applied in an energy range where the recoil effect is negligible, as it is the case for \XMM bandpass.} We also included in this modeling  the absorption associated to the interstellar medium in our own Galaxy with the WABS component. The value of the galactic column density in the direction of Mrk 766 has been fixed to $N_H=1.8\times 10^{20}\,cm^{-2}$ \cite{DL90}. An example of this fit is shown on the upper panel of Fig. \ref{fig:spectrum}. In the $0.2-10$ keV energy range both the 'soft-excess' below $\sim1$ keV and the power-law are clearly seen. We notice that the model is remarkably different from the data only at $\sim0.7$ keV and in the range $6-8$ keV, while on the rest of the $0.2-10$ keV band the description of the observed photon distribution is quite good, with $\chi^2_\nu=1112.06/1086$ in the specific case plotted in Fig. \ref{fig:spectrum}. The feature around $\sim0.7$ keV seems to be ubiquitous in all the time intervals and we tried to account for it adding an absorption edge with the EDGE component. The nature of this absorption edge in the range $\sim0.7-0.74$ keV, observed also in MCG-6-30-15 \cite{M03}, is an open issue and it has been widely debated in the literature. \cite{L96} identified an absorption feature at $\sim0.74$ keV consistent with an absorption K-edge of $O_{VII}$. On the other hand, \cite{P01} claimed that at the $red-shift$ of Mrk 766 ($z=0.0129$, \cite{WF93}), the K-edge of $O_{VII}$ would be found at $0.73$ keV, but they found no evidence of such a feature in the Reflection Grating Spectrometer (RGS) on  board \XMM. Rather, they found an absorption feature at $\sim0.7$ keV and they stated that if this is an $O_{VII}$ K-edge, it would originate in an absorbing material $red-shifted$ by more than $10000\,km\,s^{-1}$, that appears to be in contrast with the current physical scenarios for a warm absorber gas, such as an outflowing wind. Last, \cite{MB03} ascribed the feature to a relativistically broadened Ly$\alpha$ emission line of the H-like $O_{VIII}$. In our sample the feature appears with a typical energy $E_{edge}=0.71^{+0.02}_{-0.01}$ keV and an optical depth within the range $0.23-0.34$, thus our analysis does not provide any further evidence to confirm or discard one of the suggested scenarios and so we limit ourselves to include the absorption edge to the model. The addition of this component improves the fit significantly, with $\Delta\chi^2\gg2.7$ (corresponding to the 90\% confidence level). As concerns the residuals in the range $6-8$ keV plotted in the upper panel of Fig. \ref{fig:spectrum}, we distinguish a small excess with respect to the model, that we ascribe to a Fe emission line. The origin of this emission line in AGNs spectra is still controversial \cite{FM05} as well as its detection in the case of Mrk 766. In fact, \cite{L96} and \cite{N97} found evidence for a broad Fe K$\alpha$ line using \textit{ASCA} data, whereas \cite{M00} found no strong evidence of Fe emission line in \textit{BeppoSAX} data. Thus, we just add a Gaussian emission line in the range $6.40-6.97$ keV when it is required to improve the quality of the fit. We also try to model the intrinsic absorption of the galaxy hosting the AGN by adding one more WABS component and leaving the column density parameter free to vary. In all the spectra in which we use this approach, the equivalent column density drops below the Galactic column density, thus the fit results completely insensitive to this parameter and we conclude that the intrinsic absorption is negligible. This sort of the phenomenological model provides a good description ($\chi^2_\nu\sim1$) of our sample. The intrinsic photon index spans the range $\Gamma\sim1.84-2.24$, indicating that the BMC power-law slope is only slightly steeper ($\Delta \Gamma\sim0.2$) than the observed power-law found in the $2-10$ keV range (this only refers to the points above the dotted line in Fig. \ref{fig:po}). The BB color temperature we obtain from the fit oscillates between $kT=7.98\times 10^{-2}$ keV and $kT=9.29\times 10^{-2}$ keV about the average value $kT=8.59\times 10^{-2}$ keV and seems to be unrelated to the variations of the photon index.\\

\subsection{Reprocessing scenario}	\label{sec:rep}

We then included in our model a Compton reflection component using the PEXRAV model \cite{MZ95} as follows. We only considered the reflected component provided by the PEXRAV model\footnote{this is achieved by fixing the parameter rel-refl to minus one.} The other parameters were left bound to their default values, but for the normalisation $N_{pex}$ that is the only extra free parameter. We fixed the value of the incident power-law to the photon index resulting from the BMC model, so that the reprocessed photons are those emerging from the CC in the BMC scenario. The addition of the PEXRAV component produces on average a significant enhancement of the quality of the fit, with $\Delta\chi^2>2.7$ (corresponding to the 90\% confidence level). With this extra spectral component in the model the average black body color temperature given by the BMC model turned to be slightly lower ($\Delta kT=5\,eV$) with respect to the previous modeling, and the power-law steeper ($\Delta\Gamma\sim0.1$). On the contrary, the BMC normalisation remained essentially constant in the whole sample.
It is worth noting that when one calls the PEXRAV model, one of the most important parameters is the covering factor $R=\Omega/2\pi$, that describes the portion of sky that is covered by the reflecting/reprocessing medium with respect to the non-thermal radiation source (CC). In our analysis we fixed it to a negative value to obtain just the reflected component, since the incident one consisted in the BMC power-law. Usually we have that for a geometrically thin accretion disk the covering does not exceed half of the sky ($\Omega=2\pi$, $R=1$). In the most diffuse picture of the AGNs structure this holds true unless we are in one of the following cases:
\begin{enumerate}[(a)]
\item the thickness of the disk increases with the distance to the centre of the system;
\item the strong gravitational field causes the photons traveling in proximity of the event horizon (not in the direction of the observer) to be deflected and impinge on the accretion disk and to be reflected \cite{FM05}.
\end{enumerate}
A somewhat different and maybe complementary approach to explain covering factors $R>1$ consists in considering the effects of downscattering in an outflowing wind. This scenario deals with the same physical processes that cause the reflection on the accretion disk, namely the Compton reflection and the photo-absorption with subsequent re-emission, but the environment is different. Optically thick ($\tau\gtrsim2$) and cold ($T\sim10^6$ K) material traveling in the outward direction from the central object at some fraction of the speed of light is likely to be responsible for the modification of the source spectrum between $\sim10$ keV and $\sim100$ keV. Detailed analytical work and Monte Carlo simulations have been carried out by \cite{TS05} and by \cite{LT07} respectively, and satisfactory results have been obtained for the microquasar Cyg X-3 and for the AGN MCG-6-30-15. Similar results have been obtained by \cite{S10}, according to which this process can originate the so-called Compton Hump observed in a number of AGNs \cite{FM05}.\\
In a sample of ten time intervals unaffected by the eclipses, we checked whether the covering factor exceeded the critical value. This was done by generating in XSPEC a diagonal response in the $1-10\,keV$ range and running the PEXRAV model with all parameters pegged to the default values, but for the photon index $\Gamma$, which was fixed to the best-fit BMC index in the corresponding slice, and the covering factor R and the PEXRAV normalisation $N_{pex}$. We then tuned R and $N_{pex}$ to obtain the same $1-10$ keV flux measured with the true (BMC+reflection) applied model. It turned out that $R>1$ in eight out of ten attempts and usually this happens when $N_{pex}\gtrsim9\times 10^{-3}\,photons\,keV^{-1}\,cm^{-2}\,s^{-1}$ at $1$ keV in the 'true' model. This holds in particular when Mrk 766 is in the brighter and softer states found in the \XMM observations. The interpretation of this behavior is quite difficult because several processes can contribute to increase the measured covering factor, as we mentioned above. Nonetheless, the energy and profile of the Fe emission line (see Fig. \ref{fig:spectrum}) could be a further evidence in favor of the presence of outflowing material. In fact, a broad, weakly red-skewed and blue-shifted Fe emission line is predicted by \cite{S08} and \cite{S10} to arise in an obscuring wind due to scattering of line photons and fluorescence recombination. This interpretation is alternative to the 'relativistic blurring' occurring in the innermost layers of the accretion disk, whose effects on the out coming radiation would be completely smeared out by the large number of scattering events. The occultation events suggest that nearby the central engine, at least on a scale of few parsec (inferred dimension of the BLR), there is some outflowing material, probably in the form of clouds or 'comets', that moves with the speed of the order of some percent of the speed of light \cite{R11}. This in principle could be the environment where the downscattering of \xr photons takes place (\cite{TS05}). Nonetheless, further analysis is required to ascribe with no doubts the behavior of the parameter R to the down scattering in outflowing wind rather than to other physical mechanisms.

\subsection{Final baseline model}	\label{sec:bas}

The last step of our spectral analysis is to apply a self-consistent absorption model to the scenario comprising the BMC model and the reflection/reprocessing mechanism. Besides the strong feature at $0.7$ keV that we described in Sect. \ref{sec:bmc}, other small residuals are evident below $1$ keV on the upper panel of Fig. \ref{fig:spectrum} that are not modeled by the first simple phenomenological approach. \cite{TM07} found evidence of an ionized absorber in their analysis of the $1-10$ keV spectrum of Mrk 766. We replace the absorption edge with the ionized absorption model ZXIPCF \cite{MT07} which, as opposed to the ABSORI component \cite{D92}, also accounts for the absorption lines besides the absorption edges. Again, we fix the value of the incident photon index to the $\Gamma$ provided by the BMC spectral component and we set the covering factor of the absorber to one. In our idea both the power-law emerging from the Compton Cloud and the photons coming from the accretion disk (either the thermal radiation or the Compton reflected component) must pass through the warm absorber before reaching the observer at infinite. From the statistical point of view, we just substitute the two free parameters of the absorption edge (energy and optical thickness) with the column density and the ionization degree of the new component. On the lower panel of Fig. \ref{fig:spectrum} the unfolded spectrum and the resulting final best-fit model are shown for slice 5 extracted during orbit 0265. The improvement of the goodness of the fit is relevant in all slices and from the residuals plot the good quality of the fit is evident. The final step of our analysis leads to an average slight increase of the normalisations of both the BMC and PEXRAV components, as a consequence of the subtraction of flux due to the absorption. In addition, the intrinsic photon index suffers a further average steepening $\Delta\Gamma\sim0.05$ and the BB color temperature increases on average by $\Delta kT=5$ eV, giving a mean value of $kT=8.83_{-0.27}^{+0.68}\times 10^{-2}$ keV, in good agreement with the results of the BB fit to Mrk 766 soft excess performed by \cite{MM94} and \cite{B01}.\\
Hence, the final modeling comprises the Galactic absorption, the BMC model, the reflection/reprocessing scenario and a warm absorber. A Gaussian Fe K$\alpha$ emission line is also included when it is  required. The best-fit values of the main parameters of our spectral analysis are collected in Table \ref{tab:abs} in the Appendix.

\section{Discussion}	\label{sec:disc}

Our general results for \xr spectral fitting are in good agreement with the ones found in the literature. Besides the few exceptions constituted by the occultation episodes in 2005, where the source is a factor of $\sim5$ fainter, the flux computed in the $2-10\,keV$ energy range is $F_{[2-10~{\rm keV}]}\sim10^{-11}$\cgs, corresponding to a luminosity of $L_{[2-10~{\rm keV}]}\sim5\times 10^{42}\,erg\/s^{-1}$, is the same presented in \cite{L96}, \cite{B01} and \cite{R11}. Similarly, in most cases we found $\Gamma_{obs}\sim1.75-2.15$, that overlaps the range presented in \cite{SP09}, and confirms the results of \cite{L96} and \cite{M00}, except for those occulted intervals in which flatter photon indexes $\Gamma_{obs}\lesssim1.55$ are observed. In our analysis we do not build a model to parametrize the occultation episodes and we thus neglect the related times intervals (see slices 13-20 in table \ref{tab:slices} in the Appendix).\\
Our final baseline model provides a satisfactory physical picture of Mrk 766 spectra and confirms the presence of a warm absorber that was strongly suggested by \cite{TM07}. In addition, the range spanned by the intrinsic photon index $\Gamma\sim 1.9-2.4$ testifies for the characteristic steepness of the NLS1 galaxies spectra \cite{K08}.\\
The sampling we performed on the total \XMM observing time allowed us to extract forty-one 'good' time intervals that describe a number of different spectral states of the source. In Fig. \ref{fig:corr} we present the $\Gamma-N_{BMC}$ correlation obtained from these intervals (data points are re-binned in thirteen bins of $N_{BMC}$ in order to underline the magnitude of the correlation): 
\begin{figure}[h!]
\begin{center}
\includegraphics[width=0.35\textwidth,angle=270]{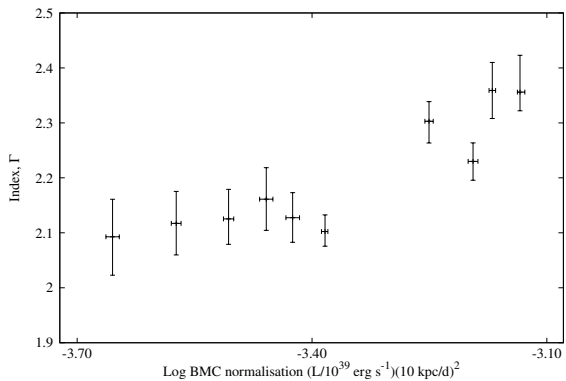}
\caption{$\Gamma-N_{BMC}$ correlation obtained excluding the eight time intervals affected by the occultation events. We re-binned in thirteen bins the $N_{BMC}$ range corresponding to the forty-one 'good' points to make the correlation between the photon index and the BMC normalisation clearer.}
\label{fig:corr}
\end{center}
\end{figure}
a positive correlation is observed, in particular a variation of the BMC normalisation of $\Delta N_{BMC}\sim6\times 10^{-4}$ implies a change of the intrinsic spectral index of $\Delta \Gamma\sim0.5$. We then argue that the small oscillations observed in the hardness-ratio light curves (see Fig. \ref{fig:lc-tot}) are driven by the same physical mechanism producing the long term spectral transitions in binary systems, namely the variations of the mass accretion rate. We recall, in fact that the luminosity, entering the definition of the normalisation of the BMC model, is proportional to $\dot{m}$, as discussed in Sect. \ref{sec:intro}. It is evident from the plot that as the normalisation increases, the spectrum becomes softer, as observed in a number of galactic \xr binaries (\cite{TS09} among others). Such a behavior was somewhat anticipated by our study of the light curves and hardness-ratio light curves. We recall that the overall increase of the luminosity occurred during orbit 0082 is accompanied by an overall spectral softening (see Fig. \ref{fig:lc-tot}) that is exactly what we would expect in the BMC scenario: as $\dot{m}$ increases the soft photon supply from the accretion disk becomes larger and larger and efficiently cools down the Compton Cloud, that shrinks. The number of up-scattered photons in the hot corona decreases and the spectrum becomes softer. The opposite occurred in orbit 0265: an overall decrease of the luminosity, manifestation of the decrease of $\dot{m}$, is related to a spectral hardening since the CC puffs out and the number of efficiently Comptonised photons increases. This correlation was in a sense anticipated by the behavior of the photon index as a function of the $2-10\,keV$ flux plotted in Fig. \ref{fig:po} and a similar correlation has been previously found for Mrk 766 by \cite{SP09} with {\it RXTE} data. Nevertheless, one must be cautious and use this relation just as an indication of this expected behavior, since the existence of a correlation has been proven only between the intrinsic photon index and the normalisation of the BMC model.\\
When we compared the correlation we found for Mrk 766 to the $\Gamma-N_{BMC}$ diagram obtained for galactic black hole sources (GBHs), we realized that what we work out for Mrk 766 is probably just a fraction of the entire spectral transition of our target (see Fig. \ref{fig:reference}). This seems to be justified by the fact that in systems powered by the mass accretion process, the physical properties, such as dimensions and time-scales, are ruled by the mass of the central object. As a consequence, given that a complete spectral evolution from a \textit{low-hard state} to a \textit{high-soft state} or vice versa for a GBH ($M_{BH}\sim10\,M_{\odot}$) lasts $\sim10-100$ days (see ST09)
%\cite{ST09}, 
the corresponding variation of the spectral index for a Super Massive Black Hole ($M_{BH}\sim10^6-10^9\,M_{\odot}$) occurs at least in $10^4$ years. In this plausible assumption, with the observations we have at hand we are essentially investigating the small spectral oscillations due to changes of $\dot{m}$ about a putative \textit{high-soft state} of Mrk 766. We also notice that the transition, towards a harder or softer state, is not smooth. In fact, in 2000 (orbit 0082) we found $N_{BMC}\sim4\times 10^{-4}\,(L/10^{39}\,erg\,s^{-1})(10\,kpc/d)^2$ and $\Gamma\sim2.15$; in 2001 (orbit 0265) the source was brighter and softer ($N_{BMC}\sim7\times 10^{-4}\,(L/10^{39}\,erg\,s^{-1})(10\,kpc/d)^2$ and $\Gamma\sim2.35$), while during 2005 (orbits from 0999 to 1004) the spectral state is on average a bit lower and harder than in 2000. Thus, the change of the spectral slope does not evolve towards a determined final state when the transition starts, but it can oscillate back and forth. As long as the parallelism between GBHs and SMBHs holds, the spectral evolution is not smooth either in binary systems. But the oscillations we detected for Mrk 766 would occur on a fraction of a second or even less, so that it would be impossible to perform an appropriate \xr spectral analysis of these oscillations.\\
Another important point to be taken into account is the magnitude of the reflected/reprocessed spectral component described by the PEXRAV model. As we mentioned before, the covering factor R increases and overcomes the limiting value $1$ when Mrk 766 is in its brighter and softer states, in particular during orbit 0265. Given that the spectrum resulting from the downscattering process is quite similar in shape to the one originating from the reflection in an ionized medium (\cite{LT07,S10}), we are tempted to say that we obtain $R>1$ when the contribution of the scattering in the outflowing wind is particularly relevant. In addition, $R>1$ in the source brighter and softer states supports this conclusion because radiation-pressure driven outflows are predicted for NLS1 galaxies accreting at large values of $\dot{m}$ \cite{PK04,K08, D12} that is likely to occur in the softer states of our target. The ionization degree of the absorbing material is not high enough (see Table \ref{tab:abs} in the Appendix) to hamper the formation of such a wind \cite{T13} whose contribution, that could constitute the ionized absorbing material described by the ZXIPCF spectral component, is strongly suggested by \cite{TM07} to account for the spectral variability of Mrk 766. The presence of this material would not only explain the spectral rising above $\sim7$ keV (and thus R>1), but it would also provide a consistent reason for the broadness of the Fe emission line and for the feature at $\sim0.7$ keV discussed in Sect. \ref{sec:bmc}. As mentioned in Sect. \ref{sec:rep}, the Fe line could be shaped by fluorescence recombination and Compton scattering in the outflowing material which cause the line to be broad and peaked at energy larger than the line rest energy \cite{S12}. As for the spectrum below $\sim1$ keV, the emission features imprinted by the outflowing wind are blended and their strength and shape are sensitive to the changes in the relative elements abundance \cite{S10}. This blending is likely to be the origin of the uncertainty on the energy of the most prominent of these features, namely the L$\alpha$ emission line of $O_{VIII}$ at $0.7$ keV. Despite this, the presence of outflowing material does not rule out the possibility that the reflection from the accretion disk contributes to the final spectrum, but with the small energy window available with \XMM no conclusive evidence can be provided in favor of one given model, because we are able to observe only the very first part of this phenomenon.\\
It is worth noting that, given the physical characteristics of the BLR clouds absorbing material as column density and ionization degree, the $0.2-0.8$ keV energy range should be the more affected by the photo-electric absorption. On the other hand, the behavior of the hardness-ratio light curves during orbit 0999 and 1000 testifies that the $2-5$ keV energy band is the most suppressed in flux. This may indicate that the partial covering of the BLR clouds mostly concerns the source of hard \xrs, rather than the source of soft \xrs. In other words, it seems plausible that the soft photons source (the accretion disk) is more extended than the hard photons source (the Compton Cloud).\\

\subsection{Scaling of the BH mass}

\begin{figure}[h!]
\begin{center}
\includegraphics[width=0.35\textwidth,angle=270]{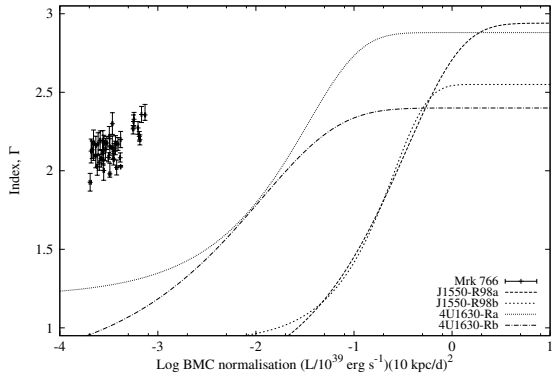}
\caption{The forty-one points of the $\Gamma-N_{BMC}$ correlation obtained for Mrk 766 are plotted along with the four reference patterns that are more likely to describe Mrk 766 (reference patterns from ST09 
%\cite{ST09} 
and \cite{ST13}).}
\label{fig:reference}
\end{center}
\end{figure}

The mass scaling technique is completely based on the shape of the $\Gamma-N_{BMC}$ correlation, that according to ST09
% \cite{ST09} 
is fit by a function
\begin{equation}	\label{eq:fit}
\Gamma(N_{BMC})=A-B\cdot \ln\left\{ \exp\left[1-\left(\frac{N_{BMC}}{N_{tr}}\right)^{\beta}\right]+1\right\}
\end{equation}
where the meaning of the parameters is the following:  coefficient A is  a value of the saturation of the intrinsic photon index, B is related to the lower value achievable by $\Gamma$, $N_{tr}$ indicates the BMC normalisation value at which $\Gamma$ starts growing and $\beta$ provides the slope of the correlation (see Fig. \ref{fig:reference}). The crucial assumption for this technique to be applied is that different sources presenting the same $\Gamma-N_{BMC}$ correlation 
%are 
undergo the same kind of spectral evolution, the only difference being the BH mass to distance squared ratio $M/d^2$ that determines the horizontal shift on the correlation plot. In particular, the appropriate reference source must be selected according to the slope of the correlation. In fact, the steepness of the correlation is tightly connected to the underlying physical process leading to the saturation of the photon index, namely it is the signature of the temperature of the converging flow. A steep slope testifies for an efficient cooling of the Compton Cloud from the soft-photon supply coming from the accretion disk and vice versa. The assumption that these two sources behave in the same way only holds if the  two correlations are as similar as possible. Thus, to scale the mass M of a target source we need to select an appropriate reference source, whose mass and distance are known, and compare its BMC normalisation N$_{BMC}$ with the one of the target at the same value of the intrinsic photon index. If the previous conditions are matched, a simple scaling relation can be worked out
\begin{equation}	\label{eq:mass}
\frac{N_t}{N_r}=\frac{M_t}{d^2_t}\frac{d^2_r}{M_r}f_G^{-1} \;\Rightarrow\; M_t=M_r\frac{N_{t}}{N_{r}}\left( \frac{d_t}{d_r}\right)^2f_G
\end{equation}
where t stands for the target, r stands for the reference and $f_G=cos\theta_r/cos\theta_t$ is a geometrical factor that depends on the respective inclination angles $\theta$ of the accretion disks with respect to the line of sight. This factor has to be considered when the accretion process is assumed to occur in disk-like geometry, while it can be neglected if spherical accretion is assumed.
Another point to be stressed is the following: usually one must compare rise (decay) transitions with rise (decay) transitions in order to be sure that the spectral evolution occurs in response to an increase (drop) of the mass accretion rate.\\
As reference sources we selected the objects presented in ST09,
%\cite{ST09}, namely GRO J1655-40, GX 339-4 and XTE J1550-564, and 4U 1630-47 presented in \cite{ST13}, 
which are all \xr binaries containing a BH of known mass and distance. Specifically, for each reference we have at hand two possible transition patterns. The available patterns for XTE J1550-564 are two rise transitions occurred in 1998, while for GRO J1655-40 we have a rise episode and a decay episode, both of them occurred in 2005. For GX 339-4 the patterns are extracted from a decay transition undergone in 2003 and from a rise transition in 2004. On the contrary, the available patterns for 4U 1630-47 are obtained from observations performed between 1996 and 2004 (in Fig. \ref{fig:reference} the label 4U 1630-Ra refers to {\it Beppo}SAX data, while 4U 1630-Rb refers to {\it RXTE} data). In Fig. \ref{fig:reference} we plotted the forty-one points extracted from Mrk 766 spectra unaffected by occultation episodes along with the four reference patterns that are more likely to describe our target source. In fact, the correlations of both GRO J1655-40 and GX 339-4 saturate at too low values of $\Gamma$, namely $\Gamma_{sat}=2.2$ and $\Gamma_{sat}=2.02$ for GRO J1655-40 rise 2005 and decay 2005, and $\Gamma_{sat}=2.08$ and $\Gamma_{sat}=2.14$ for GX 339-4 decay 2003 and rise 2004 respectively. This means that the final temperature of the Compton Cloud for these two sources is larger than for Mrk 766. In principle some problem can arise since we do not know whether the $\Gamma-N_{BMC}$ correlation for Mrk 766 testifies of a rise or a decay transition. In any case, some educated guess can solve this issue. We should  point out that even if we do not see a complete spectral evolution for the target, from 2000 to 2001 Mrk 766 spectrum underwent a softening from $\Gamma\sim2.13$ to $\Gamma\sim2.35$, while from 2001 to 2005 the spectrum essentially hardened back to almost the initial value. Then it seems that the rise and decay patterns of our target are quite similar. According to this remark we fit function (\ref{eq:fit}) to Mrk 766 points with the QDP 'ftool'\footnote{http://heasarc.gsfc.nasa.gov/docs/software/ftools/others/qdp/node3.html}. Unfortunately, since  the spectral sample only covers a small fraction of the correlation pattern we could not leave all the four parameters of the function free to vary. The only way to get a converging fit was to leave one single parameter free and constrain all the others to the values of the parameters of the reference patterns. We thus fixed the upper and lower saturation levels and the slope of the correlation to the values A, B and $\beta$ of the four references respectively. In addition, in order to better constrain the parameters, we fit function (\ref{eq:fit}) to the re-binned points presented in Fig. \ref{fig:corr}, for which the oscillation of both $\Gamma$ and $N_{BMC}$ are less pronounced. The results of this procedure are provided in Table \ref{tab:param}, where it is clear, from the value of $\chi^2_{\nu}$ that the best reference to be compared with Mrk 766 is 4U 1630-Rb.
\begin{table}[h!]
\caption{Parameters of the fit performed on Mrk 766 points.}
\begin{center}
\begin{tabular}{cccccc}
\hline
\hline
reference & A & B & $N_{tr}$ [$\times 10^{-4}$] & $\beta$ & $\chi^2$/dof \\
\hline
XTE J1550-R98a & $2.86$ & $1.50$ & $1.83^{+0.10}_{-0.10}$ & $0.50$ & $33.58/9$ \\
XTE J1550-R98b & $2.55$ & $1.21$ & $2.06^{+0.10}_{-0.08}$ & $1.00$ & $93.39/9$\\
4U 1630-Ra & $2.88$ & $1.29$ & $2.89_{-0.15}^{+0.15}$ & $0.64$ & $39.17/9$ \\ 
4U 1630-Rb & $2.40$ & $1.29$ & $0.42_{-0.04}^{+0.05}$ & $0.43$ & $16.75/9$ \\ 
\hline
\end{tabular}
\end{center}
\label{tab:param}
\end{table}

\begin{figure*}[t!]
\begin{center}
\includegraphics[width=0.4\textwidth,angle=270]{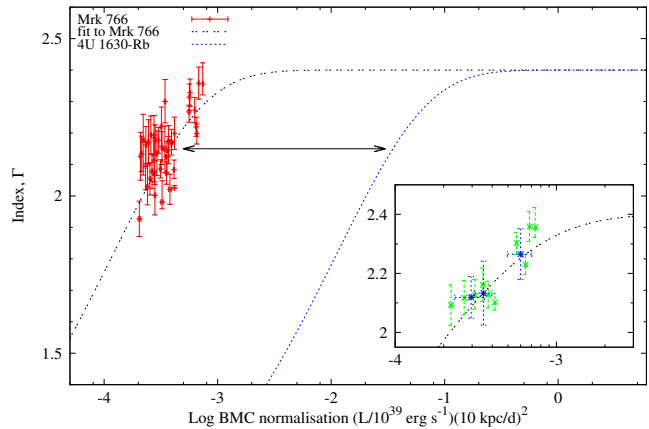}
\caption{The comparison between the $\Gamma-N_{BMC}$ correlations for Mrk 766 and for 4U 1630-Rb is shown. The black arrow stresses the fact that the two sources seem to behave the same way, the only difference being the gap in the BMC normalisation, due to the different value of the $M_{BH}/d^2$ ratio. In the small box the re-binned points (green) and the average points computed from 2000, 2001 and 2005 observations (blue) are plotted.}
\label{fig:scal}
\end{center}
\end{figure*}

In Fig. \ref{fig:scal} we plot Mrk 766 data points with its best fit curve along with the 4U 1630-Rb reference pattern: the black arrow stresses the horizontal shift due to the different mass to distance squared ratio. The small box contains the re-binned points (green) and the average values of the $\Gamma-N_{BMC}$ correlation computed in 2000, 2001 and 2005 respectively (blue). Mrk 766 seems to be properly described by the selected reference pattern and the relatively large $\chi^2$ value ($\chi^2/dof=16.75/9$) is likely to be due to the oscillations around the average values and to the small range of the correlation covered by our data points.\\
Once we select the suitable reference we can proceed with the estimate of the BH mass of Mrk 766 with formula (\ref{eq:mass}) as follows (see also Fig. \ref{fig:scal}). The mass of 4U 1630-47 is estimated to be $M_r=9.5\pm1.1\,M_{\odot}$ by \cite{ST13}. Then, as far as the distances are concerned, we use$d_r=10\,kpc$ for 4U 1630-47 \cite{ST13}, while for our target we choose $d_t=57.0\pm4.0\,Mpc$ provided by NED\footnote{http://ned.ipac.caltech.edu/}. The most uncertain term of the previous relation is the geometrical factor $f_G$. For the reference source we use $\theta_r=67\pm8^{\circ}$, but for Mrk 766 this parameter is not known with suitable precision. \cite{TM06} derived $\theta\sim29^{\circ}$ from the analysis of the Fe K$\alpha$ line profile. Furthermore, the type 1 activity shown by Mrk 766 points towards an almost face-on situation. Hence, to account for the lack of a precise measure of the inclination angle, we perform the scaling of the BH mass for $\theta_t\leq30^{\circ}$. We used function (\ref{eq:fit}) with the previously mentioned parameters to compute the BMC normalisation of the reference source $N_{r}$ for $\Gamma=1.9-2.4$, i.e. the values of the intrinsic photon index that we find for Mrk 766. Then we used formula (\ref{eq:mass}) for each point and we average over the sample, getting one evaluation of the mass for each value of the inclination angle $\theta_t$. The scaling technique that we apply provides an estimate of the central Black Hole mass for Mrk 766 of $M_{BH}=1.26^{+0.76}_{-0.76}\times 10^6\,M_{\odot}$ for $\theta_t=15^{\circ}$, to be compared with $M_{BH}=1.76^{+1.56}_{-1.40}\times 10^6\,M_{\odot}$ computed with the Reverberation Mapping method by \cite{B09}. The uncertainty of the inclination angle results in an upper limit of $M_{BH}=1.41^{+0.85}_{-0.85}\times 10^6\,M_{\odot}$ for $\theta_t=30^{\circ}$, and in a lower limit of $M_{BH}=1.22^{+0.73}_{-0.73}\times 10^6\,M_{\odot}$ for a face-on situation. The entire confidence range we obtain for the mass of Mrk 766 is then $M_{BH}=1.26^{+1.00}_{-0.77}\times 10^6\,M_{\odot}$, where the larger contributions to the uncertainty are given by the errors on the accretion disks inclination angles (33\% and 15\% given by $\theta_r$ and $\theta_t$ respectively).
Besides the issue concerning the uncertainty, the BH mass estimates performed with the scaling technique and the reverberation mapping method are in good agreement. The fact that our measure is in the lower part of the confidence range found by \cite{B09} confirms what is usually expected for Narrow Line Seyfert 1 Galaxies, i.e. relatively small BH masses and large values of the mass accretion rate. In particular, using the bolometric correction $k_{[2-10~ {\rm keV}]}=38.8_{-4.1}^{+5.4}$ presented in \cite{V09}, the average luminosity measured in our sample $L_{[2-10~{\rm keV}]}\sim5 \times 10^{42}\,erg\,s^{-1}$ and the Eddington luminosity $L_E=1.26\times 10^{38}\,\left(\frac{M_{BH}}{M_{\odot}} \right)\,erg\,s^{-1}$ we obtain $L_{bol}\sim1.94\times 10^{44}\,erg\,s^{-1}$ and $\lambda=L_{bol}/L_E\approx1$, that is slightly smaller than $\lambda\sim1.5$ estimated by \cite{VF03}. This value of $\lambda$ support the hypothesis that in the circum-nuclear regions of Mrk 766 a strong radiation-pressure driven outflowing wind rises in response to the increase of the mass accretion rate \cite{D12}.\\
From Fig. \ref{fig:scal} we can also infer that the saturation level of the photon index for Mrk 766, signature of the establishment of the full bulk-motion onto a BH, is $\Gamma\sim2.4$. Unfortunately, the \XMM observations do not provide any point on the \textit{plateau} that would conclusively prove the index saturation. Among the other \xr missions which observed Mrk 766 since 1992, ROSAT measured $F_{[0.1-2.4~{\rm keV}]}=1.5\times 10^{-10}$\cgs during the ROSAT All Sky Survey \cite{MM93}. This flux value is larger by a factor of $\sim3$ than the highest $0.1-2.4$ keV flux we obtained in our sample ($F_{[0.1-2.4~{\rm keV}]}=5.25\times 10^{-11}$\cgs, corresponding to $F_{[0.2-10~{\rm keV}]}=7.32\times 10^{-11}$\cgs from slice 5, see Table \ref{tab:abs} in the Appendix\footnote{the $0.1-2.4$ keV flux has been extrapolated using the \textit{dummyrsp} command in XSPEC.}), but ROSAT neither provides the sufficient photon statistics nor it is endowed with the suitable energy window to perform such an analysis and measure with a good precision the intrinsic photon index $\Gamma$. Nevertheless, this single ROSAT run testifies that Mrk 766 can reach even higher luminosity states than observed by \XMM. In addition, the source proved to be strongly variable in the \xr band on a time-scale of $100\,ks$ (see orbit 0265 in Fig. \ref{fig:lc-tot}) so that a dedicated campaign of pointed observations providing at least $\sim10^4$ counts per each observation could in principle find the index saturation for Mrk 766.

\subsection{The soft excess issue}

\begin{figure}[h!]
\begin{center}
\includegraphics[width=0.35\textwidth,angle=270]{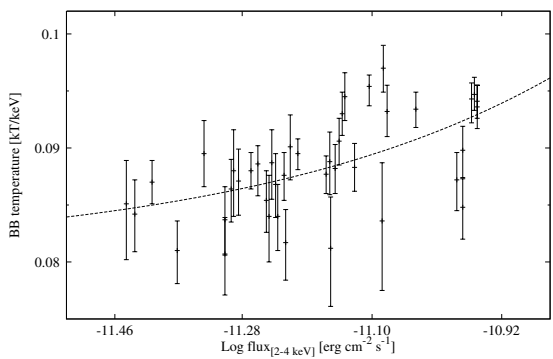}
\caption{BB color temperature (in keV) plotted against the flux in the $2-4$ keV band. The dashed line represents the linear behaviour between the two quantities. The Pearson correlation coefficient $r=0.57$ indicates a strong linear correlation.}
\label{fig:kT}
\end{center}
\end{figure}

The definition of soft excess corresponds to the increase of flux measured above the underlying power-law continuum at $E\lesssim1\,keV$ that is usually observed in AGNs spectra. On one hand the nature of this feature can be ascribed to the thermal emission from the innermost layers of the accretion disk that emit a modified BB spectrum (\cite{D12} among others). Basic and simple arguments show that the BB temperature is proportional to $M^{-1/4}$. If we assume that most of the disk luminosity is a perfect BB emission coming from within a distance $r=r_*R_g$ from the centre, where $r_*\sim2-5$ is the dimensionless radius and $R_g=GM/c^2$ is the gravitational radius, according to the Stefan-Boltzmann law we get
\begin{equation}	\label{eq:SB}
L\cong4\pi r^2\sigma_bT^4
\end{equation}
where $\sigma_b=5.67\times 10^{-5}\,erg\,cm^{-2}\,s^{-1}\,K^{-4}$ is the Stefan-Boltzmann constant and $2\pi r^2$ is the surface of the two-sided disk. Hence, provided that the luminosity equals some fraction $\lambda$ of the Eddington luminosity $L_E$, the expected disk temperature turns out to be
\begin{equation}	\label{eq:T} 
T_{exp}\cong\left( \frac{\lambda L_E}{4\pi r^2 \sigma_b} \right)^{1/4} \propto M^{-1/4}\,\lambda^{1/4}.
\end{equation}
As far Mrk 766 is concerned, this physical scenario is plausible. In fact, performing an order of magnitude estimate by considering the central value of our confidence range for the BH mass in Mrk 766, the previously estimated $\lambda\approx1$ and $r_*\sim6-7$, we obtain $kT_{exp}\sim4.7-7.5\,\times 10^{-2}$ keV, that partially overlaps the range spanned by the BB colour temperature provided by our baseline model $kT_{fit}=8.83_{-0.27}^{+0.68}\times 10^{-2}$ keV (see Sect. \ref{sec:bas}).\\
Unfortunately for this interpretation, some problem arises comparing $kT_{exp}$ with the disk temperature $kT_{fit}$ resulting from a BB spectral fit of the soft excess for AGNs hosting BHs whose mass is $\sim10^7-10^9\,M_\odot$: the expected value spans the range $kT_{exp}\sim10-40$ eV, whereas the usual fit value is $kT_{fit}\sim0.1$ keV.\\
On the other hand, the origin of the soft excess can be explained by \xr Compton reflection and photo-ionisation on the accretion disk \cite{MZ95}, sometimes combined with relativistic blurring of emission lines below $1\,keV$ \cite{FM05}. The combination of these two phenomena could produce a bump above the underlying power-law. \cite{MB03} claimed that this very scenario produces good results when applied to Mrk 766. The recent discovery of soft/negative time lags in Mrk 766 spectra by \cite{E11} and \cite{De13} seems to support the reverberation scenario, as the variations in the soft \xr band ($0.3-0.7$ keV) are driven by the variations in the hard \xr band ($1.5-4$ keV). Nevertheless, this only holds on relatively short time-scales ($t\sim1\,ks$), whereas it is clear from the previous papers that for longer characteristic time-scales ($t\gtrsim5\,ks$) the soft band drives the changes in the hard band. This is consistent with the scenario presented in this paper. In fact, the short time-scales soft/negative lags would be justified by the energy released in the accretion disk by the \xr radiation: energetic photons traveling in the inward direction lose energy in the dense medium via photo-absorption and Compton recoil. Basically, they deposit energy in the accretion disk, indeed increasing its temperature. This energy is then re-emitted, mainly in the UV and soft \xr bands. This mechanism was suggested by \cite{BS74} to explain the emission from the surface of a normal star in a binary system, but it seems plausible that it works also in this situation. To strengthen this conclusion, in Fig. \ref{fig:kT} we plot the BB disk temperature against the flux in the $2-4$ keV band, for which we computed the Pearson correlation coefficient $r=0.57$ which denotes a strong linear correlation between the two quantities.
In principle, this mechanism also explains the difference between $kT_{exp}$ and $kT_{fit}$ that we obtain for Mrk 766. In fact, the simple estimate we performed using equation (\ref{eq:T}) does not account for the contribution to the disk temperature given by the energy deposited by the \xr radiation.\\
Thus, in this plausible scenario, the short-term variations ($t\sim1\,ks$, \cite{De13}) in the soft \xr band, which are dictated by the impingement of high energy photons on the accretion disk, determine the large spread of the points in the $\Gamma-N_{BMC}$ correlation (see red points in Fig. \ref{fig:scal}). On the other hand, the long-term variations are driven by the soft \xr band, which in turn is ruled by the propagation of the mass accretion rate through the disk \cite{L97}. This gives rise to the correlation plotted in Fig. \ref{fig:scal} (specifically green and blue points).\\
\\

As a future development of the present work, we aim to extend this analysis to other NLS1 galaxies, to check whether the correlation we find for Mrk 766 between the photon index $\Gamma$ and the normalisation of the BMC model $N_{BMC}$ holds for other AGNs with $M_{BH}\sim10^6\,M_{\odot}$. Furthermore, the extension of the sample of spectral transitions, both for GBHs and AGNs, would be the key to understand whether there is a finite or infinite number of possible transition patterns, which could provide some constrain on the physics at work in accreting compact objects.\\

\subsection{Alternative AGN spectral models}
This section is devoted to a very brief review of some alternative models used in the literature to describe the spectral variability which is the main focus of this paper. For a complete discussion we address the interested reader to the papers we quoted and to the references therein.\\
\cite{SP09}, hereafter SP09, also find the correlation of the photon index  $\Gamma$ with the dimensionless mass accretion rate $\dot m$ combining the results for different AGN sources. In particular, they demonstrate this correlation using a phenomenological model, power-law plus line plus edge  applying {\it RXTE} data  for a number of  the AGN  sources (Mkr 766, NGC 3227, NGC 5548 NGC 5506 and NGC 3516). It is worth noting that SP09 define $\dot m$ as ratio of the average flux in the energy range from 2 to 10 keV, $F_{2-10~{\rm keV}}$ to the Eddington luminosity $L_{Ed}$ which is $1.3\times 10^{38}M/M_{\odot}$ erg s$^{-1}$ for an AGN with a BH mass $M$. Although the  flux from the disk as a source of the soft photons  in an AGN should be calculated in the energy range much lower than 2 keV, see  a typical disk temperature  and its dependence on the energy flux in 2-4 keV range in Fig.  \ref{fig:kT}. SP09 emphasize that the index vs mass accretion correlation reflects a {\it true/intrinsic} correlation between the photon index of the power-law component and accretion rate. This statement is similar to that which we claim in our presented paper.   SP09 also point out  that it is widely believed that hard X-rays from AGN are produced by the {\it thermal} Comptonisation.
 SP09 claim that the enhancement factor $\eta _{\rm comp}$ due to thermal Componization depends on the geometry of the accretion flow while  
 %\cite{SP09}
 \cite{st85}   demonstrate that $\eta _{\rm comp}$  is determined by the properties of the hot plasma and seed photons, namely the spectral index $\alpha=\Gamma-1$, the plasma temperature $kT_e$ and the seed  photon temperature $kT_s$.  SP09 also suggest the index vs mass accretion, $\dot m$ correlation can be explained if  $\eta _{\rm comp}$ is proportional to $\dot m$.  
Finally, SP09 conclude that the ``observed ''$\Gamma-F_{2-10~{\rm keV}}$''  or $\Gamma-\dot m$, can be explained using the reflection effect with  the constant  reflection amplitude $R=\Omega/2\pi=1$, where $\Omega$ is the solid angle constant  covered by the cold material as viewed from the X-ray source if one can assume that the power-law continuum varies in flux and shape.\\
\cite{TM07}, hereafter T07, investigate the origin of the high variability of Mrk 766 with two different models, both of them relying on a power-law in the range $1-10\,keV$ that is constant in slope and variable in normalisation. In the first model a constant scattered component and an ionised reflection component are assumed to play an important role and the observed spectrum is the result of the relative strength of directly viewed and reflected/scattered components. The second model mainly ascribes the spectral variability to the presence of complex layers of absorbing material partially covering the cxentral source of radiation. These layers of gas, perhaps arising from an outflowing wind, are free to vary both in covering fraction, ionisation degree and column density. T07 claim that the most robust description of the spectral variability is given in terms of directly viewed and scattered or absorbed fractions of flux, even though it is difficult to understand whether the continuum drives the variations of the absorbing material or whether the continuum is intrinsically constant and the observed variability is simply a consequence of the changes in the covering fraction. In addition, the degeneracy between the ionisation degree and the column density makes impossible to understand which of these parameters is actually responsible for the spectral changes.\\
Another physical process often addressed to explain the spectral variability in AGNs is Compton reflection from ionised or partially ionised material (see for example \cite{FM05}). As already mentioned in Sect. \ref{sec:rep}, according to this scenario the putative accretion disk or the dusty torus are illuminated by the hard X-rays giving rise to a 'reflection' spectrum dominated by fluorescent K$\alpha$ line from the most abundant elements, particularly iron. One possible drawback of this model is the fact that sometimes the fraction of sky $R=\Omega/2\pi$ occupied by the accretion disk with respect to the source of X-rays results to be larger than the maximum expected value $R=1$, as we discussed in the previous section. The issue can be solved taking into account general relativistic effects taking place in proximity of the central BH. In fact, depending on the height of the hard X-ray source above the disk \cite{MF04}, the light bending phenomenon deflects photon that would travel to infinite causing them to be intercepted by the disk, increasing \textit{de facto} the factor $R$, or to fall into the hole event horizon.

\subsection{An effect of outflow  on the emergent spectra}
\cite{tsa07} showed that the outflow can be launched from the accretion disk if the local mass accretion rate $\dot M_{loc}$ is  higher than the Eddington one.  The disk works a filter that does not allow to supply $\dot M_{loc}$ higher than a certain critical  value. The resulting  Thomson optical optical depth  of the outflow 
$\tau_W$ can be higher than 1  using  typical parameters of the  wind and disk. 
\cite{tkb} studied an effect of outflow on the emergent spectra from compact objects (NS and BH).
They demonstrated solving analytically  the radiative Fokker-Planck  equation that the emergent iron line profile formed in the outflow of the  optical depth   of order of 1 expanding with the outflow velocity of 0.05-0.1 of the speed of light $c$   leads  to the formation of a broad red-shifted and skewed line feature.  
Later this result was confirmed by \cite{LT07}  who used Monte Carlo simulations to investigate 
 the iron line profile formed in the outflow.  On the other hand, using XMM-{\it Newton} monitoring
 of Mkr 766,   \cite{TM07} and \cite{R11} found an outflow component of the velocity spanning from 0.01 to 0.05c which can lead to the formation of  the broad red-skewed iron lines observed in many Galactic and extragalactic sources (see e.g. a review by \cite{m07}). 

 %Independently of these aforementioned  studies 
   
%\cite{RM06}.
\section{Conclusions}	\label{sec:conc}

We  study the timing, spectral and accretion properties of the NLS1 galaxy Mrk 766 by exploiting an intense \XMM monitoring from May 2000 to June 2005, for an overall observing time of $\sim711\,ks$ ($\sim430\,ks$ of effective monitoring). We study the light curves and the hardness-ratio light curves to isolate time intervals corresponding to different spectral states and study the resulting spectral transition pattern. This led to the selection of forty-nine time slices lasting at least $3\,ks$ and containing at least $3\times 10^4$ photons where the spectra were extracted. From the time resolved spectral analysis it emerges that Mrk 766 spectrum is satisfactorily described by a model comprising the Galactic absorption, a simple Comptonisation model (bulk-motion Comptonisation model BMC currently used for GBHs), a reflection/reprocessing component, and a warm absorber. A Gaussian iron emission line is also included when statistically required. The average $2-10$ keV measured flux is $F_{[2-10~
{\rm keV}]}\sim10^{-11}$\cgs, corresponding to a luminosity of $L_{[2-10~{\rm keV}]}\sim5\times 10^{42}\,erg\,s^{-1}$. Twenty per cent of the whole observing time is probably affected by BLR clouds occultations occurring during the observations performed in 2005 \cite{R11}, for which the just mentioned spectral description was not completely physically reliable. Hence, we neglected the time intervals related to the eclipses.\\
We observed luminosity increases lasting few thousands seconds that proved to be related to general intrinsic softening of the spectrum, as well as luminosity drops connected to spectral hardening in the photon index, in agreement with the theoretical expectations. The pattern of the correlation between the slope and the normalisation of the model we work out testifies of spectral changes about a putative \textit{high-soft state} of Mrk 766 rather than a real spectral transition like those observed for GBHs. This seems to be consistent with the fact that for large systems like AGNs the expected time-scales for a complete spectral transition ($\sim10^4$ years) are much longer than for stellar BHs ($\sim10-100$ days). Nevertheless, the shape of the correlation is a conclusive evidence that those small spectral changes are indeed driven by the same physical process that causes the complete spectral evolutions seen in \xr binary systems, namely the variations of the mass accretion rate $\dot{m}$. We used the reference $\Gamma-N_{BMC}$ correlation of the GBH 4U 1630-47, for which both mass and distance are known, to derive the mass of Mrk 766. The obtained value of $M_{BH}=1.26^{+1.00}_{-0.77}\times 10^6\,M_{\odot}$ is in perfect agreement with the mass estimate performed with the reverberation mapping method, proving that this scaling technique is a powerful and reliable tool to estimate the mass of SMBHs in AGNs, provided the suitable quality of \xr data and a moderate knowledge of the inclination angle of the accretion disk with respect to the line of sight.\\
In addition, the mass estimate allowed us to compare the disk color temperature provided by our baseline model ($kT_{fit}=8.83_{-0.27}^{+0.68}\times 10^{-2}\,keV$) with the one computed with a simple model involving a prefect BB emission from the innermost regions of the accretion disk ($kT_{\exp}\sim4.7-7.5\,\times 10^{-2}\,keV$), that is slightly smaller but still comparable to the best fit value. This would also justify the discovery of soft/negative time lags on $t\sim1\,ks$ time-scales.\\
The comparison of our target diagram with the reference source also pointed out that the possible saturation level of the intrinsic photon index for Mrk 766 is at $\Gamma\sim2.4$. The index saturation would provide an important piece of evidence on the nature of the compact object sitting in the centre of AGNs. Such a measure is within reach of a dedicated observations campaign to Mrk 766.

\section*{Acknowledgements}

We thank Guido Risaliti and Chris Done for stimulating discussions and suggestions and Piero Ranalli for his support in the installation of the SAS tool. Elena Seifina is gratefully acknowledged for providing fundamental data on 4U 1630-47. SG would like to thank the IMPRS for Astronomy and Cosmic Physics for the financial support. RG acknowledges support from the Italian Space Agency (ASI) under the contract ASI-INAF I/009/10/0.

\appendix

\section{Appendix: details of spectral analysis}

We provide the table collecting the fifty time intervals extracted from the \XMM observations of Mrk 766 from May 2000 to June 2005. For each time slice the initial and final time (expressed as detector time), the net exposure and the total amount of counts are given. Furthermore, we present the best-fit values of the main parameters of the applied baseline model, along with the values of the reduced $\chi^2$ stating the goodness of the fit. The eight time intervals affected by the occultation events (slices 13 through 20) and slice 47 have been neglected.
\begin{table*}[b]
\caption{Results of the spectral sampling.}
\begin{center}
\resizebox*{0.9\textwidth}{!}{
\begin{tabular}{cccccc|cccccc}
\hline
\hline
Slice & $t_{start}^a$ & $t_{stop}^a$ & net & total & orbit & Time & $t_{start}^a$ & $t_{stop}^a$ & net & total & orbit \\
 & & & exposure$^b$ & counts & & slice & & & exposure$^b$ & counts & \\
\hline
$1$ & $0.75207$ & $0.75220$ & $8.57$ & $134019$ & $0082$ & $26$ & $2.33605$ & $2.33615$ & $6.59$ & $89374$ & $1001$  \\ 
$2$ & $0.75220$ & $0.75232$ & $8.41$ & $125636$ & $0082$ & $27$ & $2.33615$ & $2.33625$ & $6.94$ & $85746$ & $1001$  \\ 
$3$ & $0.75232$ & $0.75244$ & $8.56$ & $170706$ & $0082$ & $28$ & $2.33625$ & $2.33634$ & $6.31$ & $89651$ & $1001$  \\ 
$4$ & $1.06736$ & $1.06747$ & $7.45$ & $245568$ & $0265$ & $29$ & $2.33634$ & $2.33643$ & $6.31$ & $105675$ & $1001$  \\ 
$5$ & $1.06747$ & $1.06760$ & $9.11$ & $326828$ & $0265$ & $30$ & $2.33643$ & $2.33659$ & $11.22$ & $204204$ & $1001$  \\ 
$6$ & $1.06760$ & $1.06773$ & $9.11$ & $289675$ & $0265$ & $31$ & $2.33659$ & $2.33675$ & $11.18$ & $152048$ & $1001$  \\ 
$7$ & $1.06773$ & $1.06787$ & $9.82$ & $266728$ & $0265$ & $32$ & $2.33675$ & $2.33684$ & $6.31$ & $70649$ & $1001$  \\ 
$8$ & $1.06787$ & $1.06800$ & $9.06$ & $283484$ & $0265$ & $33$ & $2.33684$ & $2.33699$ & $10.91$ & $139757$ & $1001$  \\ 
$9$ & $1.06800$ & $1.06813$ & $9.12$ & $288279$ & $0265$ & $34$ & $2.33781$ & $2.33786$ & $3.11$ & $58107$ & $1002$  \\ 
$10$ & $1.06813$ & $1.06827$ & $9.61$ & $277075$ & $0265$ & $35$ & $2.33786$ & $2.33810$ & $16.77$ & $256078$ & $1002$  \\ 
$11$ & $1.06827$ & $1.06840$ & $9.12$ & $267828$ & $0265$ & $36$ & $2.33810$ & $2.33822$ & $8.41$ & $164053$ & $1002$  \\ 
$12$ & $1.06840$ & $1.06862$ & $15.77$ & $267617$ & $0265$ & $37$ & $2.33822$ & $2.33834$ & $8.41$ & $150947$ & $1002$  \\ 
$13^c$ & $2.33263$ & $2.33283$ & $13.50$ & $62289$ & $0999$ & $38$ & $2.33834$ & $2.33845$ & $7.68$ & $143367$ & $1002$  \\ 
$14^c$ & $2.33283$ & $2.33292$ & $6.31$ & $30118$ & $0999$ & $39$ & $2.33845$ & $2.33855$ & $7.01$ & $139289$ & $1002$  \\ 
$15^c$ & $2.33292$ & $2.33301$ & $10.54$ & $63629$ & $0999$ & $40$ & $2.33855$ & $2.33866$ & $6.59$ & $134918$ & $1002$  \\ 
$16^c$ & $2.33301$ & $2.33322$ & $10.52$ & $76954$ & $0999$ & $41$ & $2.33866$ & $2.33875$ & $6.54$ & $92120$ & $1002$  \\ 
$17^c$ & $2.33322$ & $2.33340$ & $12.64$ & $57031$ & $0999$ & $42$ & $2.33951$ & $2.33965$ & $9.68$ & $180264$ & $1003$  \\ 
$18^c$ & $2.33340$ & $2.33366$ & $11.76$ & $46840$ & $0999$ & $43$ & $2.33965$ & $2.33978$ & $9.11$ & $152568$ & $1003$  \\
$19^c$ & $2.33433$ & $2.33440$ & $5.16$ & $30153$ & $1000$ & $44$ & $2.33978$ & $2.34000$ & $15.42$ & $246411$ & $1003$  \\ 
$20^c$ & $2.33440$ & $2.33452$ & $8.44$ & $41207$ & $1000$ & $45$ & $2.34000$ & $2.34020$ & $14.01$ & $190676$ & $1003$  \\ 
$21$ & $2.33452$ & $2.33467$ & $10.53$ & $135942$ & $1000$ & $46$ & $2.34020$ & $2.34034$ & $9.81$ & $94548$ & $1003$  \\ 
$22$ & $2.33467$ & $2.33479$ & $8.41$ & $102879$ & $1000$ & $47^d$ & $2.34034$ & $2.34050$ & $2.59$ & $19094$ & $1003$ \\ 
$23$ & $2.33479$ & $2.33494$ & $10.51$ & $151659$ & $1000$ & $48$ & $2.34187$ & $2.34193$ & $4.36$ & $44136$ & $1004$  \\ 
$24$ & $2.33494$ & $2.33512$ & $12.58$ & $130706$ & $1000$ & $49$ & $2.34193$ & $2.34208$ & $10.46$ & $124369$ & $1004$  \\ 
$25$ & $2.33512$ & $2.33524$ & $8.41$ & $83773$ & $1000$ & $50$ & $2.34208$ & $2.34217$ & $5.27$ & $84421$ & $1004$ \\ 
\hline
\end{tabular}}
\end{center}
\caption*{The net exposure accounts for the detector live time and the filtering from flaring background. The values of $t_{start}$ and $t_{stop}$ are measured from the begenning of the \XMM mission.\\
$^a$ In units of $10^8\,s$.\\
$^b$ In units of $10^3\,s$.\\
$^c$ Affected by occultation episodes.\\
$^d$ Excluded because not matching of the sampling criteria.}
\label{tab:slices}
\end{table*}

\begin{table*}[h]
\caption{Best-fit parameters of the baseline applied model.}
\begin{center}
\resizebox*{0.9\textwidth}{!}{
\begin{tabular}{ccccccccc}
\hline
\hline
Slice & $kT^a$ & $N_{bmc}^b$ & $\Gamma^c$ & $N_{pex}^d$ & $N_H^e$ & $\xi^f$ & $F^g$ & $\chi^2$/dof \\
\hline
$1$ & $8.17_{-0.33}^{+0.29}$ & $3.20_{-0.04}^{+0.04}$ & $2.22_{-0.06}^{+0.06}$ & $3.68_{-1.05}^{+1.25}$ & $0.20_{-0.05}^{+0.06}$ & $20.71_{}^{}    $ & $3.62$ & $916.63$/$908$ \\ 
$2$ & $8.86_{-0.28}^{+0.16}$ & $3.24_{-0.08}^{+0.08}$ & $1.98_{-0.02}^{+0.07}$ & $0.07_{-0.07}^{+0.97}$ & $0.09_{-0.02}^{+0.07}$ & $10.02_{-2.34}^{+4.78}$ & $3.35$ & $947.22$/$848$ \\ 
$3$ & $8.82_{-0.22}^{+0.21}$ & $4.17_{-0.05}^{+0.04}$ & $2.20_{-0.05}^{+0.05}$ & $1.88_{-0.83}^{+0.96}$ & $0.19_{-0.04}^{+0.05}$ & $24.18_{-3.66}^{+3.18}$ & $4.34$ & $998.94$/$924$ \\ 
$4$ & $8.98_{-0.24}^{+0.21}$ & $6.81_{-0.07}^{+0.06}$ & $2.36_{-0.05}^{+0.05}$ & $6.15_{-1.68}^{+1.93}$ & $0.17_{-0.03}^{+0.04}$ & $17.49_{-3.33}^{+2.28}$ & $6.89$ & $1055.78$/$1020$ \\ 
$5$ & $9.43_{-0.21}^{+0.14}$ & $7.39_{-0.05}^{+0.11}$ & $2.36_{-0.03}^{+0.07}$ & $4.83_{-1.48}^{+2.67}$ & $0.13_{-0.02}^{+0.04}$ & $12.32_{-0.96}^{+2.18}$ & $7.32$ & $1189.27$/$1109$ \\ 
$6$ & $9.47_{-0.14}^{+0.15}$ & $6.55_{-0.11}^{+0.11}$ & $2.20_{-0.03}^{+0.03}$ & $1.51_{-1.25}^{+0.98}$ & $0.11_{-0.01}^{+0.02}$ & $8.47_{-0.92}^{+0.86} $ & $6.78$ & $1264.31$/$1115$ \\ 
$7$ & $9.34_{-0.16}^{+0.15}$ & $5.60_{-0.10}^{+0.10}$ & $2.27_{-0.03}^{+0.02}$ & $3.19_{-0.89}^{+0.98}$ & $0.13_{-0.02}^{+0.02}$ & $9.38_{-0.98}^{+3.47} $ & $5.74$ & $1153.85$/$1069$ \\ 
$8$ & $9.36_{-0.19}^{+0.19}$ & $6.28_{-0.06}^{+0.06}$ & $2.27_{-0.04}^{+0.04}$ & $3.25_{-1.13}^{+1.26}$ & $0.20_{-0.03}^{+0.03}$ & $17.72_{-2.06}^{+1.85}$ & $6.71$ & $1151.00$/$1105$ \\ 
$9$ & $9.41_{-0.15}^{+0.14}$ & $6.47_{-0.11}^{+0.10}$ & $2.22_{-0.03}^{+0.03}$ & $2.27_{-0.86}^{+0.95}$ & $0.12_{-0.01}^{+0.02}$ & $8.11_{-0.85}^{+0.80} $ & $6.77$ & $1141.04$/$1118$ \\ 
$10$ & $8.48_{-0.28}^{+0.25}$ & $5.64_{-0.06}^{+0.05}$ & $2.31_{-0.04}^{+0.04}$ & $5.73_{-1.33}^{+1.49}$ & $0.23_{-0.04}^{+0.04}$ & $23.81_{-2.55}^{+2.23}$ & $6.36$ & $1201.71$/$1140$ \\ 
$11$ & $8.72_{-0.27}^{+0.24}$ & $5.71_{-0.06}^{+0.05}$ & $2.33_{-0.04}^{+0.04}$ & $3.63_{-1.43}^{+1.53}$ & $0.20_{-0.03}^{+0.04}$ & $21.55_{-2.51}^{+2.20}$ & $6.29$ & $1127.48$/$1085$ \\ 
$12$ & $8.95_{-0.14}^{+0.13}$ & $3.54_{-0.06}^{+0.06}$ & $2.07_{-0.04}^{+0.04}$ & $0.58_{-0.56}^{+0.63}$ & $0.09_{-0.01}^{+0.02}$ & $8.79_{-1.30}^{+1.17} $ & $3.78$ & $1342.81$/$1234$ \\ 
$21$ & $8.54_{-0.28}^{+0.26}$ & $2.53_{-0.03}^{+0.03}$ & $2.05_{-0.05}^{+0.05}$ & $1.31_{-0.57}^{+0.65}$ & $0.29_{-0.06}^{+0.08}$ & $32.88_{-4.19}^{+3.69}$ & $3.11$ & $984.13$/$959$ \\ 
$22$ & $8.64_{-0.29}^{+0.26}$ & $2.41_{-0.04}^{+0.03}$ & $2.03_{-0.06}^{+0.07}$ & $0.69_{-0.50}^{+0.71}$ & $0.18_{-0.05}^{+0.06}$ & $27.15_{-5.41}^{+4.71}$ & $2.86$ & $903.21$/$814$ \\ 
$23$ & $8.40_{-0.30}^{+0.28}$ & $2.82_{-0.04}^{+0.03}$ & $2.18_{-0.05}^{+0.05}$ & $1.80_{-0.67}^{+0.78}$ & $0.28_{-0.06}^{+0.07}$ & $30.54_{-3.71}^{+3.24}$ & $3.32$ & $981.42$/$952$ \\ 
$24$ & $8.10_{-0.29}^{+0.26}$ & $2.14_{-0.03}^{+0.02}$ & $2.13_{-0.06}^{+0.06}$ & $1.77_{-0.56}^{+0.66}$ & $0.25_{-0.06}^{+0.07}$ & $26.29_{-3.69}^{+3.17}$ & $2.46$ & $902.52$/$913$ \\ 
$25$ & $8.42_{-0.33}^{+0.30}$ & $2.09_{-0.03}^{+0.04}$ & $2.13_{-0.08}^{+0.08}$ & $1.48_{-0.63}^{+0.78}$ & $0.22_{-0.07}^{+0.10}$ & $33.19_{-7.27}^{+5.77}$ & $2.27$ & $757.83$/$740$ \\ 
\hline 
\end{tabular}}
\end{center}
\label{tab:abs}
\end{table*}

\begin{table*}[h]
\caption*{Table \ref{tab:abs} continues from previous page}
\begin{center}
\resizebox*{0.9\textwidth}{!}{
\begin{tabular}{ccccccccc}
\hline
\hline
Slice & $kT^a$ & $N_{bmc}^b$ & $\Gamma^c$ & $N_{pex}^d$ & $N_H^e$ & $\xi^f$ & $F^g$ & $\chi^2$/dof \\
\hline
$26$ & $8.71_{-0.30}^{+0.28}$ & $2.76_{-0.05}^{+0.04}$ & $2.11_{-0.07}^{+0.07}$ & $1.66_{-0.77}^{+0.94}$ & $0.21_{-0.06}^{+0.08}$ & $30.37_{-6.10}^{+5.08}$ & $3.08$ & $815.43 $/$747$ \\ 
$27$ & $8.80_{-0.40}^{+0.36}$ & $2.46_{-0.04}^{+0.04}$ & $2.17_{-0.07}^{+0.07}$ & $2.16_{-0.86}^{+1.05}$ & $0.26_{-0.08}^{+0.08}$ & $18.83_{-5.21}^{+3.00}$ & $2.89$ & $773.01 $/$758$ \\ 
$28$ & $8.40_{-0.40}^{+0.36}$ & $2.75_{-0.04}^{+0.04}$ & $2.19_{-0.06}^{+0.06}$ & $1.78_{-0.83}^{+0.99}$ & $0.23_{-0.06}^{+0.08}$ & $22.41_{-4.29}^{+3.49}$ & $3.24$ & $743.77 $/$755$ \\ 
$29$ & $8.87_{-0.32}^{+0.29}$ & $3.44_{-0.05}^{+0.05}$ & $2.30_{-0.07}^{+0.07}$ & $3.60_{-1.21}^{+1.48}$ & $0.24_{-0.06}^{+0.07}$ & $26.96_{-4.33}^{+3.71}$ & $3.63$ & $709.54 $/$764$ \\ 
$30$ & $9.06_{-0.21}^{+0.20}$ & $3.69_{-0.04}^{+0.04}$ & $2.15_{-0.04}^{+0.04}$ & $1.29_{-0.66}^{+0.75}$ & $0.26_{-0.04}^{+0.05}$ & $26.07_{-2.91}^{+2.62}$ & $4.10$ & $1054.23$/$1051$ \\ 
$31$ & $8.68_{-0.29}^{+0.27}$ & $2.64_{-0.04}^{+0.03}$ & $2.08_{-0.05}^{+0.05}$ & $1.45_{-0.64}^{+0.74}$ & $0.21_{-0.04}^{+0.06}$ & $21.45_{-3.65}^{+3.04}$ & $3.24$ & $1021.00$/$996$ \\ 
$32$ & $8.95_{-0.29}^{+0.29}$ & $2.32_{-0.08}^{+0.08}$ & $2.09_{-0.07}^{+0.08}$ & $1.03_{-0.72}^{+0.88}$ & $0.15_{-0.04}^{+0.03}$ & $6.92_{-1.48}^{+1.15} $ & $2.61$ & $710.66 $/$706$ \\ 
$33$ & $8.37_{-0.31}^{+0.29}$ & $2.58_{-0.03}^{+0.03}$ & $2.19_{-0.06}^{+0.06}$ & $2.01_{-0.80}^{+0.92}$ & $0.25_{-0.06}^{+0.07}$ & $26.59_{-26.59}^{}    $ & $2.92$ & $1206.77$/$1192$ \\ 
$34$ & $8.36_{-0.61}^{+0.51}$ & $3.43_{-0.07}^{+0.08}$ & $2.15_{-0.07}^{+0.07}$ & $2.00_{-1.28}^{+1.56}$ & $0.28_{-0.08}^{+0.12}$ & $26.09_{-5.23}^{+4.21}$ & $4.40$ & $608.27 $/$661$ \\ 
$35$ & $8.76_{-0.22}^{+0.20}$ & $3.02_{-0.03}^{+0.03}$ & $2.18_{-0.04}^{+0.04}$ & $2.02_{-0.57}^{+0.65}$ & $0.21_{-0.03}^{+0.04}$ & $20.68_{-2.54}^{+2.23}$ & $3.46$ & $1127.82$/$1150$ \\ 
$36$ & $9.30_{-0.19}^{+0.19}$ & $3.92_{-0.08}^{+0.09}$ & $2.17_{-0.04}^{+0.04}$ & $0.95_{-0.82}^{+0.89}$ & $0.12_{-0.02}^{+0.03}$ & $7.82_{-1.05}^{+0.97} $ & $4.26$ & $935.55 $/$923$ \\ 
$37$ & $9.45_{-0.21}^{+0.21}$ & $3.57_{-0.08}^{+0.10}$ & $2.12_{-0.05}^{+0.05}$ & $1.49_{-0.72}^{+0.83}$ & $0.15_{-0.03}^{+0.02}$ & $8.50_{-0.96}^{+0.94} $ & $4.08$ & $943.91 $/$948$ \\ 
$38$ & $9.70_{-0.21}^{+0.20}$ & $3.80_{-0.10}^{+0.11}$ & $2.02_{-0.05}^{+0.05}$ & $0.90_{-0.67}^{+0.77}$ & $0.16_{-0.02}^{+0.02}$ & $7.12_{-0.79}^{+0.78} $ & $4.42$ & $994.43 $/$956$ \\ 
$39$ & $9.54_{-0.17}^{+0.10}$ & $4.17_{-0.02}^{+0.03}$ & $2.03_{-0.00}^{+0.01}$ & $0.07_{-0.07}^{}     $ & $0.32_{-0.03}^{+0.03}$ & $31.24_{-2.80}^{+1.90}$ & $4.48$ & $974.43 $/$902$ \\ 
$40$ & $9.32_{-0.22}^{+0.23}$ & $4.13_{-0.06}^{+0.03}$ & $2.08_{-0.03}^{+0.03}$ & $0.30_{-0.30}^{+0.57}$ & $0.34_{-0.06}^{+0.09}$ & $32.83_{-3.88}^{+3.66}$ & $4.65$ & $977.13 $/$915$ \\ 
$41$ & $9.01_{-0.29}^{+0.28}$ & $2.80_{-0.04}^{+0.05}$ & $2.00_{-0.06}^{+0.06}$ & $1.36_{-0.69}^{+0.83}$ & $0.26_{-0.06}^{+0.09}$ & $28.97_{-5.12}^{+4.36}$ & $3.40$ & $879.46 $/$809$ \\ 
$42$ & $8.83_{-0.21}^{+0.21}$ & $3.70_{-0.07}^{+0.08}$ & $2.17_{-0.05}^{+0.05}$ & $1.78_{-0.81}^{+1.55}$ & $0.15_{-0.02}^{+0.02}$ & $9.33_{-0.99}^{+0.88} $ & $4.26$ & $985.30 $/$1000$ \\ 
$43$ & $8.88_{-0.29}^{+0.26}$ & $3.24_{-0.04}^{+0.04}$ & $2.15_{-0.05}^{+0.05}$ & $2.37_{-0.82}^{+0.94}$ & $0.19_{-0.04}^{+0.05}$ & $17.08_{-3.56}^{+2.44}$ & $3.91$ & $955.51 $/$959$ \\ 
$44$ & $8.77_{-0.17}^{+0.16}$ & $3.14_{-0.05}^{+0.06}$ & $2.09_{-0.03}^{+0.03}$ & $0.90_{-0.42}^{+0.46}$ & $0.12_{-0.02}^{+0.02}$ & $7.07_{-0.79}^{+0.73} $ & $3.75$ & $1170.01$/$1163$ \\ 
$45$ & $8.80_{-0.16}^{+0.16}$ & $2.77_{-0.06}^{+0.05}$ & $2.07_{-0.04}^{+0.04}$ & $0.44_{-0.44}^{+0.51}$ & $0.11_{-0.02}^{+0.02}$ & $8.64_{-1.23}^{}      $ & $3.10$ & $1055.64$/$1045$ \\ 
$46$ & $8.70_{-0.19}^{+0.19}$ & $2.04_{-0.05}^{+0.06}$ & $1.93_{-0.06}^{+0.06}$ & $0.67_{-0.41}^{+0.48}$ & $0.07_{-0.07}^{+0.02}$ & $5.45_{-2.22}^{+1.64} $ & $2.32$ & $824.40 $/$800$ \\ 
$48$ & $8.51_{-0.49}^{+0.38}$ & $2.21_{-0.05}^{+0.05}$ & $2.18_{-0.11}^{+0.08}$ & $2.48_{-1.11}^{+1.53}$ & $0.27_{-0.10}^{+0.20}$ & $33.82_{-9.39}^{+7.14}$ & $2.31$ & $591.83 $/$567$ \\ 
$49$ & $8.07_{-0.36}^{+0.32}$ & $2.33_{-0.03}^{+0.03}$ & $2.16_{-0.06}^{+0.06}$ & $2.41_{-0.73}^{+0.85}$ & $0.23_{-0.05}^{+0.07}$ & $24.39_{-3.79}^{+3.17}$ & $2.83$ & $948.39 $/$901$ \\ 
$50$ & $8.12_{-0.51}^{+0.45}$ & $2.92_{-0.04}^{+0.05}$ & $2.14_{-0.07}^{+0.07}$ & $2.06_{-1.06}^{+1.27}$ & $0.18_{-0.09}^{+0.07}$ & $19.19_{-10.44}^{+3.55}$ & $3.80$ & $812.27$/$830$ \\ 
\hline

\end{tabular}}
\end{center}
\caption*{The missing time intervals (13-20,47) refer to the occultation episodes which have been neglected.\\
$^a$ BB colour temperature in unit s of $10^{-2}\,keV$.\\
$^b$ Normalisation of the BMC component in units of $10^{-4}\,\left(\frac{L}{10^{39}\,erg\,s^{-1}}\right)\times \left(\frac{10\,kpc}{d}\right)^2$.\\
$^c$ Intrinsic photon index.\\
$^d$ Normalisation of the PEXRAV component in units of $10^{-2}\,photons\,keV^{-1}\,cm^{-2}\,s^{-1}$ at $1\,keV$.\\
$^e$ Column density of the ZXIPCF component in units of $10^{22}\,cm^{-2}$.\\
$^f$ Ionisation degree of the ZXIPCF component in units of $erg\,cm\,s^{-1}$.\\
$^g$ $0.2-10\,keV$ flux in units of $10^{-11}$\cgs.}
\end{table*}

\bibliographystyle{plain} % style aa.bst
\bibliography{biblio} % your references Yourfile.bib

\end{document}